\documentclass[aps,pre,reprint,groupedaddress, showkeys]{revtex4-2}
\usepackage{physics}
\usepackage{siunitx}
\usepackage{mathtools}
\usepackage{amssymb}
\usepackage{bm}      
\usepackage{graphicx}
\graphicspath{{./Figures/}}
\usepackage{hyperref}
\hypersetup{unicode=true, colorlinks=true, linkcolor=blue, citecolor=blue}
\usepackage{titlesec}
\titlespacing*{\subsection}{0pt}{*3}{0pt} 

\usepackage{enumitem}
\usepackage{textcase}
\makeatletter

\def\rtx@sectionbase{%
  \@startsection{section}{1}{\z@}%
    {1\baselineskip}
    {1\baselineskip}
    {\normalfont\bfseries\centering}
}

\renewcommand{\section}{\@ifstar{\rtx@sectstar}{\rtx@sect}}
\newcommand{\rtx@sect}[2][]{%
  \rtx@sectionbase[\MakeTextUppercase{#1}]{\MakeTextUppercase{#2}}%
}
\newcommand{\rtx@sectstar}[1]{%
  \rtx@sectionbase*{\MakeTextUppercase{#1}}%
}

\makeatother

\begin{document}

\preprint{APS/123-QED}

\title{Maximum-Entropy Model of Colored Noise in Superdiffusive Axonal Growth} 

\author{Julian Sutaria}
\affiliation{Department of Physics and Astronomy, Tufts University, Medford, MA, 02155, USA}
\author{Cristian Staii}%
\email{cstaii01@tufts.edu}
\affiliation{Department of Physics and Astronomy, Tufts University, Medford, MA, 02155, USA}

\date{\today}

\begin{abstract}
We develop a coarse-grained stochastic theory for axonal growth on micropatterned substrates using the Shannon--Jaynes maximum entropy principle. Starting from a Langevin description of growth cone motion, we infer the effective distribution of traction force relaxation rates from experimentally motivated constraints rather than postulating the colored noise directly. The resulting relaxation rate distribution generates a stationary colored acceleration process with power-law temporal correlations and yields analytical predictions for the axonal mean squared displacement and velocity autocorrelation. The long-time behavior is controlled by the slow-relaxation part of the inferred distribution, corresponding physically to broadly distributed clutch or adhesion engagement times. For biologically relevant parameters, the model predicts a negative correlation exponent $\alpha=-1/2$. This prediction is in close quantitative agreement with measurements on cortical neurons cultured on micropatterned poly-D-lysine-coated PDMS substrates, which are well described by $\alpha\simeq -0.6$ and exhibit superdiffusive mean squared displacement scaling with exponent $1.4$. The same framework accounts for the crossover from early diffusive behavior to long-time anomalous growth and for the corresponding power law decay of the velocity autocorrelation. These results show how entropy-constrained active fluctuations can connect microscopic force generation processes to emergent growth laws in neuronal systems and, more broadly, in active matter.
\end{abstract}

\maketitle

\section{Introduction}

Neurons are the fundamental information-processing units of the brain, consisting of a cell body, branched dendrites that receive signals, and long axons that transmit impulses to other cells \cite{Franze2010, Lowery2009, Huber2003}. Understanding how axons grow, respond to external cues, and self-organize into functional networks remains a central problem at the interface of biophysics, developmental neuroscience, and nonequilibrium statistical mechanics \cite{DeGennes2007, Franze2010, Alert2020, Oliveri2022}. Axonal extension is a multiscale process in which molecular motors, actin and microtubule dynamics, adhesion turnover, membrane tension, and environmental geometry collectively shape the motion of the growth cone \cite{Franze2020, DeRooij2018, Coles2015, Athamneh2015}. Because these processes interact across widely separated spatial and temporal scales, a major challenge is to identify a reduced set of variables and principles that can account for the experimentally observed growth statistics without requiring a complete bottom-up reconstruction of the mechanochemistry \cite{Alert2020, Oliveri2022, Duswald2024}.

A wide range of theoretical approaches has been used to study neuronal growth, including diffusion-based descriptions, biased random walks, continuum mechanics, mechanochemical feedback models, and molecular-level treatments of cytoskeletal force generation \cite{Rizzo2013, Vensi2019, Descoteaux2022, DeRooij2017, DeRooij2018, Jakobs2020, Oliveri2021}. In addition, numerous experiments have established that geometrical cues, adhesion, and internal force generation all play important roles in controlling neuronal extension and guidance \cite{Franze2020, Koch2012, Lilja2018, breau2023chemical, pillai2026longrange}. At the same time, the correlation properties and temporal organization of the fluctuations that drive axonal advance remains less well understood. In many biological systems, including cell migration and collective cellular motion, coarse-grained dynamics are often modeled with effective stochastic equations whose noise terms are modeled phenomenologically rather than inferred from experimentally accessible constraints \cite{Alert2020, Dieterich2008, amselem2012stochastic, Duswald2024}. This leaves open a basic question: how should one construct a stochastic theory of axonal growth when the microscopic force-generating processes are only partially known, yet the long-time transport properties are directly measurable?

\begin{figure*}
     \centering
    \includegraphics[width=\textwidth]{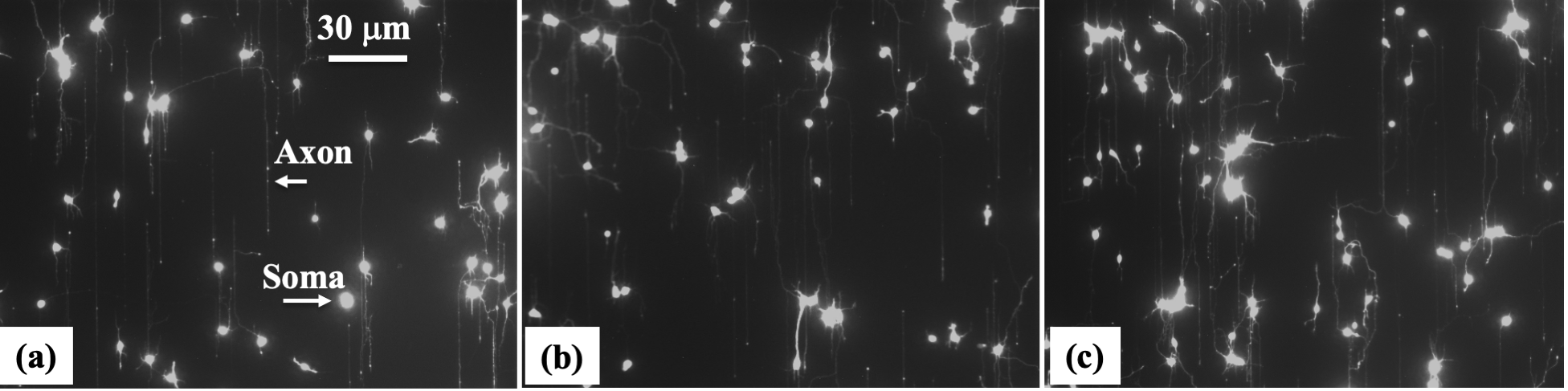}
    \caption{Fluorescence images of cultured cortical neurons on micropatterned PDL-coated PDMS substrates with spatial period $d=3~\mu\mathrm{m}$, showing strong axonal alignment along the direction of the micropatterns. Images are shown at (a) $t=70$ hrs, (b) $t=80$ hrs, and (c) $t=90$ hrs, after plating. The pattern spatial period $d$ is defined in Fig.~\ref{fig2}. The main structural components of a neuronal cell are labeled in (a), and the scale bar in panel (a) applies to all images. In the analysis presented here, we track the longest neurite, treated as the axon, to quantify population-level growth statistics on micropatterned substrates.}    
    \label{fig1}
 \end{figure*}

\begin{figure*}[t!]
     \centering
    \includegraphics[width=0.8\linewidth]{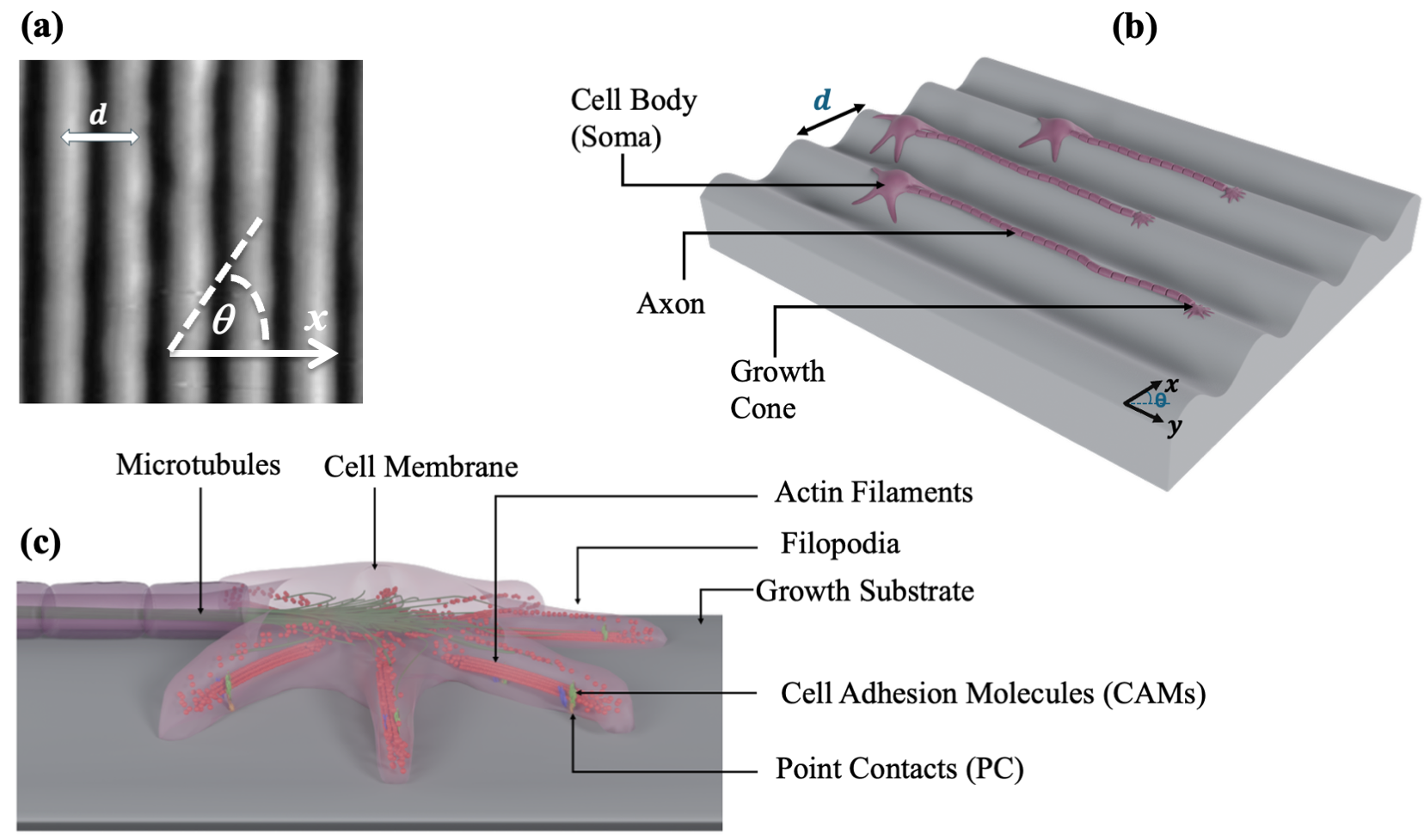}
    \caption{(a) Topographic atomic force microscope (AFM) image of a micropatterned PDMS growth substrate ($d=3\mu m$). (b) Schematic representation of the micropatterned PDMS surface, with periodic grooves oriented along the $y$ direction and spatial period $d$ along $x$. Cortical neurons extend a long axon aligned with the grooves and several shorter dendrites. The growth angle, $\theta(t)$, is defined as the angle between the axonal velocity and the $x$-axis at time $t$. (c) Schematic illustration of the growth cone and clutch mechanism, highlighting key cytoskeletal components. Interactions between cell adhesion molecules (integrins, cadherins) and actin filaments generate traction forces that drive forward the motion of the growth cone.  Details on the fabrication of PDMS substrates and neuron culture are provided in the main text.}    
    \label{fig2}
 \end{figure*}

Engineered micropatterned substrates provide a particularly useful setting in which to address this question. By imposing well-defined geometric anisotropy, they make it possible to probe the relation between cell-substrate interactions and axonal transport in a controlled and reproducible way \cite{kundu2013superimposed, vecchi2024geometry, Vensi2019, Descoteaux2022, Kumarasinghe2022, gladkov2017design, sands2024interface}. For example, in previous work \cite{Yurchenko2019}, we demonstrated that axons growing along periodic microfabricated patterns exhibit superdiffusive motion, with mean squared displacement (MSD) scaling as $\langle\Delta x^{2}(t)\rangle\propto t^{\nu}$ over time scales of several tens of hours, where $\nu\simeq 1.4$. These results were obtained using cortical neurons cultured on poly-D-lysine-coated polydimethyl-siloxane (PDMS) substrates patterned with periodic parallel ridges separated by grooves \cite{Vensi2019, Descoteaux2022, Yurchenko2019, Basso2019} (Fig.~\ref{fig1} and ~\ref{fig2}). The same experiments also revealed long-lived velocity correlations and a clear crossover from an early diffusive regime to a later anomalous-growth regime \cite{Yurchenko2019}. While existing mechanochemical models describe important processes such as actin polymerization, molecular clutch binding, and substrate interactions, they do not by themselves provide a direct statistical route from measurable microscopic constraints to the observed power-law scaling behavior and long-time transport exponents \cite{Franze2020, Oliveri2022, DeRooij2018}.

This gap motivates the central idea of the present work: when a fully bottom-up mechanochemical theory is difficult to formulate at the level needed for direct comparison with measured growth statistics, one can instead seek a controlled effective description by combining experimentally accessible constraints with a principle of statistical inference. In statistical physics, the Shannon--Jaynes maximum-entropy (MaxEnt) principle provides precisely such a framework: given partial information, it selects the least-biased probability distribution consistent with that information \cite{PhysRev.106.620, RevModPhys.85.1115}. MaxEnt and related information-theoretic constructions are common in equilibrium and nonequilibrium physics, where they are used to infer effective ensembles, rate distributions, and mesoscopic laws from incomplete information \cite{PhysRev.106.620, RevModPhys.85.1115}. By contrast, in biological physics they are rarely used to derive stochastic transport laws directly from measurable dynamical constraints, especially in systems where the fluctuations themselves carry the signature of active force generation. Here we develop such an information-theoretic construction for axonal growth and use it to derive an effective transport description directly from experimentally accessible constraints.

In this paper we show that a constrained Shannon--Jaynes MaxEnt framework yields a predictive stochastic theory of axonal growth. Starting from a coarse-grained Langevin description of growth cone motion and imposing only experimentally motivated constraints on the protrusive acceleration noise---stationarity, zero mean, and a finite mean adhesion lifetime---we demonstrate that the MaxEnt principle selects a \textit{unique} probability distribution consistent with these assumptions. Applying MaxEnt to a superposition of Ornstein--Uhlenbeck (OU) traction pulses yields a Gamma-distributed spectrum, $p(\lambda)\propto \lambda^{-(1+\alpha)} e^{-\beta\lambda}$, which produces a colored acceleration noise with autocorrelation $C_{a}(\tau)\propto |\tau|^{\alpha}$, where $\alpha<0$. For biologically relevant parameters, the model predicts $\alpha=-1/2$, quantitatively reproducing the observed superdiffusive scaling without adjustable parameters and revealing heavy-tailed force fluctuations as a direct consequence of minimum information constraints.

A central feature of this approach is that the expression for the colored noise fluctuations driving axonal extension is not postulated \emph{a priori}, but is instead inferred from a constrained maximum-entropy problem. In this way, the paper introduces an information theoretic framework for biological growth fluctuations that connects partial experimental knowledge to a predictive stochastic equation of motion. The significance of this framework is both system--specific and general. At the level of axonal growth, the theory yields quantitative predictions for the mean squared displacement and the velocity autocorrelation on micropatterned substrates, and links the observed superdiffusive exponents to the statistics of traction-force fluctuations. In particular, it selects the biologically relevant exponent $\alpha=-1/2$ from the heavy-tailed lifetime statistics associated with active biopolymer disassembly and clutch-mediated traction transmission, thereby providing a direct mechanistic interpretation of the measured power laws. At a broader level, the framework places axonal growth alongside other driven systems that display anomalous transport and long-memory dynamics, including active colloids, cell migration, L\'evy-like search processes, and other nonequilibrium systems with long-memory transport \cite{Peruani2007, Dieterich2008, Tilles2017, Reynolds2012, Metzler2000}. More generally, it provides a concrete example of how entropy-based maximization principles can be used in biological physics to construct controlled effective theories from partial data when detailed microscale modeling is underdetermined.

Our analysis also identifies experimentally testable consequences. Because the long-time dynamics are governed by the inferred spectrum of traction-force relaxation rates, perturbations that alter adhesion turnover, actomyosin activity, or cytoskeletal stability should modify the noise amplitude, crossover time, or asymptotic scaling exponent. This opens the way to direct experimental tests through pharmacological modulation of clutch unbinding, ATP depletion, or perturbations of actin and microtubule dynamics using compounds such as blebbistatin or taxol, combined with the fluorescence-tracking protocols already established for cortical neurons on micropatterned PDMS \cite{Descoteaux2022, Basso2019, Kumarasinghe2022}. The theory therefore not only reproduces an observed exponent but also provides a broader set of predictions that can be used to discriminate between competing microscopic scenarios.

This paper is organized as follows. In \hyperref[sec:experiment]{Section~II} we present details of experimental procedure and data analysis. In \hyperref[sec:clutch]{Section~III} we introduce the molecular-clutch picture and the effective Langevin description for axonal extension on micropatterned substrates. In \hyperref[sec:MaxEnt]{Section~IV} we formulate the maximum-entropy problem and derive the effective rate spectrum for traction-force fluctuations. In \hyperref[sec:Super]{Section~V} we show how this spectrum generates long-memory acceleration correlations and leads to explicit predictions for the axonal mean squared displacement and velocity autocorrelation. We then compare these predictions with experimental measurements for cortical neurons on PDMS micropatterns and discuss their robustness and physical interpretation. \hyperref[sec:Discussion]{Section~VI} provides a discussion of the broader implications of entropy-constrained noise for axonal growth and for active biological transport more generally. \hyperref[sec:Conclusions]{Section~VII} summarizes the main results.

 \begin{figure*}
     \centering
    \includegraphics[width=0.8\linewidth]{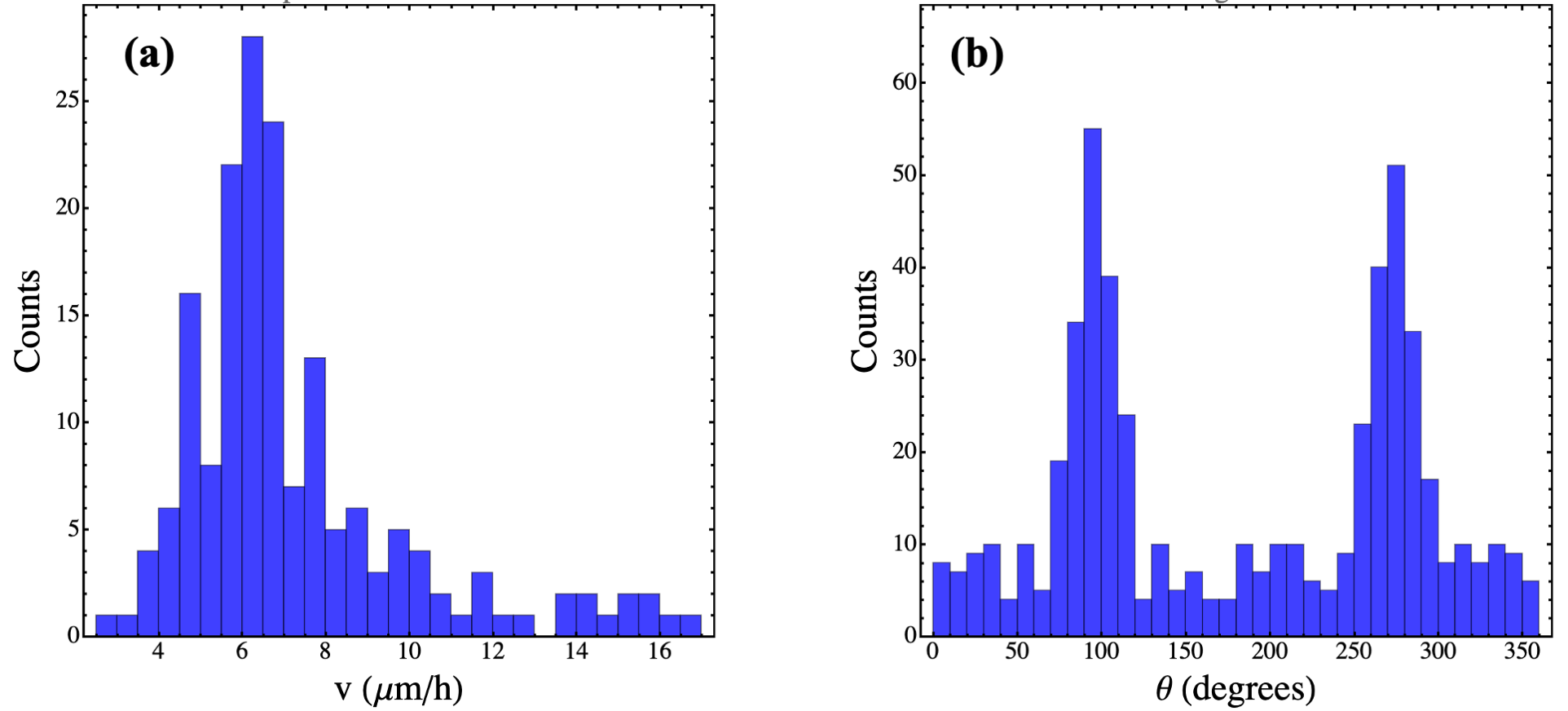}
   \caption{Examples of the growth cone speed distribution (a) and axonal growth angle distribution (b) for neurons cultured on micropatterned PDMS substrates with pattern spatial period $d=3~\mu\mathrm{m}$. The data were obtained at $t=60~\mathrm{hrs}$ after neuron plating. The angular distribution in (b) shows strong axonal alignment with the surface patterns, with pronounced peaks at $\theta=90^\circ$ and $270^\circ$.}   
    \label{fig3}
 \end{figure*}

\section{Experimental Details}\label{sec:experiment}

\textit{Neuronal Cell Culture}. Primary cortical neurons were isolated from embryonic day 18 rat embryos. All procedures involving animal tissue were approved by the Tufts University Institutional Animal Care and Use Committee and were carried out in accordance with NIH guidelines for the care and use of laboratory animals. Neuronal dissociation and culture were performed according to established protocols described in our previous work \cite{Kumarasinghe2022, Yurchenko2019, Vensi2019, Basso2019, Descoteaux2022}. First, cortices were incubated in 5~mL of trypsin at 37\textdegree{}C for 20~min. Enzymatic digestion was then quenched by adding 10~mL of soybean trypsin inhibitor (Life Technologies). The tissue was mechanically dissociated, centrifuged, and the supernatant was removed. The resulting pellet was resuspended in 20~mL of Neurobasal medium (Life Technologies) supplemented with GlutaMAX, b27 (Life Technologies), and penicillin-streptomycin. For fluorescence imaging, live cortical samples were rinsed once with phosphate-buffered saline (PBS) and incubated at 37\textdegree{}C for 30~min in PBS containing 50~nM Tubulin Tracker Green (Oregon Green 488 Taxol, bis-Acetate, Life Technologies, Grand Island, NY). After incubation, samples were rinsed twice with PBS and transferred to fresh PBS for imaging. Fluorescence images were acquired using a standard FITC filter set (excitation/emission: 495~nm/521~nm), and axon outgrowth was quantified with the NeuronJ plugin in ImageJ. Immunostaining measurements reported in earlier studies confirmed the high neuronal purity of these cultures. Cells were plated onto micropatterned polydimethylsiloxane (PDMS) substrates pre-coated with poly-D-lysine (PDL; 0.1~mg/mL, Sigma-Aldrich, St. Louis, MO) at a density of 4,000~cells/cm$^2$. As shown previously, neuronal cultures maintained at low densities (3,000--7,000~cells/cm$^2$) favor the development of long axons and are therefore well suited for studies of axonal dynamics under controlled surface cues \cite{Kumarasinghe2022, Yurchenko2019, Vensi2019, Basso2019, Descoteaux2022}. Examples of neuronal growth are shown in Fig.~\ref{fig1}.

\textit{Micropatterned Substrates}. The PDMS substrates contained parallel ridges separated by grooves (Fig.~\ref{fig2}). These micropatterns were fabricated by a simple imprinting procedure in which diffraction gratings were pressed into uncured PDMS, thereby generating periodic surface features with spatial period $d$. To ensure uniform surface functionalization, the PDL coating was applied by spin coating. Additional details of substrate preparation and micropatterning are given in Refs.~\cite{Kumarasinghe2022, Yurchenko2019, Vensi2019, Basso2019, Descoteaux2022}.

\textit{Imaging and Data Acquisition}. Atomic force microscopy (AFM) and fluorescence imaging were used to characterize both the substrates and neuronal growth. AFM images were acquired with an MFP-3D system (Asylum Research) equipped with a BioHeater fluid cell and integrated with an inverted Nikon Eclipse Ti microscope (Micro Video Instruments, Avon, MA). Fluorescence images of neurons were obtained using standard FITC filters (excitation: 495~nm; emission: 521~nm).

\textit{Data Analysis}. Growth cone dynamics were analyzed in ImageJ (NIH). Growth cone positions were tracked from fluorescence images acquired every $\Delta t = 5$~min over a total observation interval of 30~min. Measurements were performed at multiple times after plating, specifically at $t=5$ and $10~\mathrm{h}$, and subsequently at 10-hour intervals up to $90~\mathrm{h}$. The choice $\Delta t = 5$~min ensured that the growth cone displacement $\Delta r$ exceeded the spatial resolution of the measurement (approximately $0.1~\mu\mathrm{m}$), while the ratio $\Delta r/\Delta t$ remained a good approximation to the instantaneous growth velocity $V$. Fig.~\ref{fig3} shows representative growth cone speed and growth angle distributions for neurons cultured on micropatterned PDMS substrates with $d=3~\mu\mathrm{m}$, obtained at $t=60~\mathrm{hrs}$ after neuron plating.

\section{Molecular clutch and topographical feedback} \label{sec:clutch}
Growth cones convert environmental signals into directed motion by coordinating cytoskeletal dynamics and cell-substrate adhesion through a molecular clutch mechanism \cite{Lowery2009, Huber2003, Franze2010, Franze2020}.
In this model, actin filaments polymerize at the leading edge (actin treadmilling), while myosin II motors pull
actin filaments together and induce retrograde flow. Transmembrane adhesion receptors such as integrins and cadherins form point contacts (PCs) with the
substrate, mechanically linking the actin network to the
extracellular environment and modulating traction forces (Fig.~\ref{fig2}(c)). Swanson and Wingreen demonstrated that active biopolymers such as actin filaments or microtubules can be modeled as ordered chains of bound subunits \cite{Swanson2011}. In the continuum limit, the distribution of polymer lifetimes exhibits a diffusive prefactor $t^{-3/2}$, imparting a heavy-tailed character to the dynamics. We propose that these long-lived filaments store elastic energy, which is intermittently released as traction forces exerted against the substrate. In addition, the micropattern topography imposes an effective linear spring with stiffness~$k_{\perp}$,
which suppresses lateral deviations while leaving longitudinal motion weakly
damped \cite{Descoteaux2022, Vensi2019}. As a result, intermittent bursts of actin and microtubule polymerisation become synchronized through substrate‑dependent
traction regulation. The stochastic nature of these underlying processes generates a colored force noise that enters the coarse‑grained Langevin description of axonal extension $x(t)$:
\begin{equation}
  \ddot x + \gamma \dot x = a(t),
  \label{eq:langevin}
\end{equation}
with $\gamma>0$ the damping constant, and $a(t)$ the net protrusive acceleration generated by cytoskeletal filaments. 

\textit{Physical origin of $a(t)$.}
In the clutch framework, traction is transmitted while adhesions are engaged: actin retrograde flow is partially arrested as nascent point contacts mature, producing forward motion of the growth cone opposite to retrograde flow \cite{Franze2009,Franze2020}. During an engaged interval of lifetime $\tau_{\rm adh}$ the effective traction builds as the clutch strengthens and then terminates upon detachment \cite{Koch2012,Hyland2014}. We coarse-grain the ensemble of engagement events into a family of modes characterized by relaxation rates $\lambda>0$. Each event excites a traction/acceleration response with its own $\lambda$, whose distribution $p(\lambda)$ will be derived from the MaxEnt principle below. As a result, we model the dynamics using a colored acceleration noise $a(t)$ that aggregates intermittent polymerization and traction events:
\begin{equation}
  a(t)=\int_{0}^{\infty}\!\eta_{\lambda}(t)\,d\lambda,\qquad
  \dot{\eta}_{\lambda}(t)=-\lambda\,\eta_{\lambda}(t)+\sqrt{2D_{\lambda}}\,\xi_{\lambda}(t),
  \label{eq:spectral}
\end{equation}
where $\xi_{\lambda}(t)$ are independent Gaussian white noises, and $D_{\lambda}$ collects mode weights (pulse energy and event rate). Eq.~\eqref{eq:spectral} is an effective overdamped relaxation (linear-response) description for force/acceleration contributions during an engaged interval, with $1/\lambda$ the corresponding correlation/termination time. Microscopically, adhesion turnover and clutch detachment provide a finite cutoff that renders single-engagement correlations short-ranged. This modal representation therefore captures a broad class of engaged-interval pulse shapes (including ramp-and-terminate pulses) without assuming that adhesion strength decays from the moment of engagement $t=0$. Under this condition, the long-time acceleration autocorrelation $C_a(\tau)$ and the resulting MSD exponent are controlled by the small–$\lambda$ behavior of $p(\lambda)$, not by the detailed shape of a single pulse \cite{SM}.

For a single mode with rate $\lambda$ (lifetime $\tau_{\rm adh}=1/\lambda$), the Ornstein–Uhlenbeck covariance is
$\langle\eta_{\lambda}(t)\eta_{\lambda}(t+\tau)\rangle=(D_{\lambda}/\lambda)\,e^{-\lambda|\tau|}$. Independence across $\lambda$ implies that covariances add linearly hence:
\begin{equation}
  C_a(\tau)\equiv\langle a(t)\,a(t+\tau)\rangle
  =\int_{0}^{\infty}\frac{D_{\lambda}}{\lambda}\,e^{-\lambda|\tau|}\,d\lambda.
  \label{eq:Ctau}
\end{equation}
We fix the normalization by writing $D_{\lambda}/\lambda=\rho_{a}^{2}\,p(\lambda)$, with $\rho_a^{2}$ the variance of the distribution. Consequently $C_a(\tau)$ is proportional to the Laplace transform of $p(\lambda)$, 
so its long-time decay is governed by the small–$\lambda$ tail of $p(\lambda)$ (see Supplemental Material, Secs.~S4 and~S5 \cite{SM}).

\section{Maximum‑entropy derivation of the probability density $p(\lambda)$}\label{sec:MaxEnt}
Building on the molecular clutch mechanism, the active biopolymer model of axonal growth, and the mechanical coupling between the axon and the micropatterned substrate,
we use the maximum‑entropy principle to derive the probability distribution $p(\lambda)$ governing
$a(t)$, and show that it naturally gives rise to power-law temporal correlations with a negative exponent.

 The Shannon–Jaynes (MaxEnt) principle states that, among all candidate probability distributions $p(\lambda)$ consistent with a given set of constraints, the distribution that maximizes the entropy \cite{PhysRev.106.620, RevModPhys.85.1115}:
\begin{equation}
  S[p]= -\int_0^{\infty} p(\lambda)\,\ln\!\frac{p(\lambda)}{m(\lambda)}\,d\lambda ,
  \label{eq:entropy}
\end{equation}
introduces the \emph{least} additional bias beyond what is imposed by the known information. Here, 
$m (\lambda)$ represents a prior measure (or reference distribution) that encodes baseline knowledge in the absence of constraints.

To determine this distribution, we maximize the entropy in Eq.~\eqref{eq:entropy} subject to two constraints: 
(1) normalization, $\int_{0}^{\infty}p(\lambda)\,d\lambda=1$;
  and (2) a finite mean relaxation rate $\langle\lambda\rangle=\Lambda_{1}\sim k_{\perp}/\zeta$, where $\Lambda_1$ is determined by the cell-substrate interactions through the effective lateral stiffness $k_{\perp}$ and the viscous damping coefficient $\zeta$, associated with motion perpendicular to the micropatterned grooves \cite{Descoteaux2022}.

Introducing Lagrange multipliers $(\chi,\beta)$ for the two
constraints,  we extremize the functional:
\begin{equation}
\mathcal{J}[p]=S[p]
       -\chi\Bigl(\int p(\lambda)\,d\lambda-1\Bigr)
       -\beta\Bigl(\int \lambda p(\lambda)\,d\lambda-\langle\lambda\rangle\Bigr), 
\end{equation}
\noindent
using a physics-informed Jeffreys prior $m(\lambda)=\lambda^{-(1+\alpha)}$. Because $\lambda$ is a positive rate and no particular time scale is preferred, we start from a scale-invariant baseline $m_0 \propto 1/\lambda$ (equal weight per decade) and modify it by the independently motivated small-$\lambda$ scaling implied by the heavy-tailed engagement lifetimes. 

Functional differentiation $\delta\mathcal{J}/\delta p=0$ yields (see Sec.~S1 of the Supplemental Material \cite{SM}):
\begin{equation}
  p(\lambda)=\frac{1}{Z}\,
             \lambda^{-(1+\alpha)}\,
             e^{-\beta\lambda},
  \qquad
  Z=\int_{0}^{\infty}\lambda^{-(1+\alpha)}e^{-\lambda\beta}\,d\lambda.
  \label{eq:plambda}
\end{equation}
\noindent
Note that the integral converges both at \(\lambda\to0^{+}\) and
\(\lambda\to\infty\) if $\beta>0$ and \(\alpha<0\). By introducing the Gamma function,
\(\displaystyle\Gamma(-\alpha)=\int_{0}^{\infty}u^{-\alpha-1}e^{-u}\,du\)
(for \(\alpha<0\)), and performing elementary Gamma-function algebra we obtain (details in Sec.~S1 of the Supplemental Material \cite{SM}):
\begin{equation}
Z=\beta^{\alpha}\,\Gamma(-\alpha).  
\end{equation}

The fixed mean \(\langle\lambda\rangle\) determines the scale for the exponential cutoff \(\beta\): $ \langle\lambda\rangle
  =\frac{1}{Z}\int_{0}^{\infty}
      \lambda^{-\alpha}\,e^{-\lambda\beta}\,d\lambda
  \;\Longrightarrow\;$
 \noindent 
$\beta=\frac{\Gamma(1-\alpha)}{\langle\lambda\rangle\,\Gamma(-\alpha)}$.

\textit{Acceleration autocorrelation at long times.}
Substituting the MaxEnt solution for $p(\lambda)$, Eq.~\eqref{eq:plambda}, into the Laplace transform expression for the acceleration autocorrelation function, Eq.~\eqref{eq:Ctau}, gives
\begin{equation}
\begin{split}
  C_a(\tau)
  &= \langle a(t)\,a(t+\tau)\rangle
   = \rho_a^2 \int_0^{\infty} p(\lambda)
      e^{-\lambda |\tau|} \, d\lambda  \\
  &= \frac{\rho_a^2}{Z}
     \int_0^{\infty} \lambda^{-(1+\alpha)}
     e^{-(\beta+|\tau|)\lambda} \, d\lambda \\
  &= \rho_a^2\,
     \frac{\Gamma(-\alpha)(\beta+|\tau|)^{\alpha}}
          {\Gamma(-\alpha)\beta^{\alpha}} .
\end{split}
\end{equation}
Here we have used the Gamma-function identity
\begin{equation}
\int_0^{\infty} \lambda^{-(1+\alpha)}
e^{-(\beta+|\tau|)\lambda} \, d\lambda
= \Gamma(-\alpha)(\beta+|\tau|)^{\alpha},
\end{equation}
which is valid for $\alpha<0$ and $\beta+|\tau|>0$. Therefore,
\begin{equation}
    C_a(\tau)
  = \rho_{a}^{2}\left(1+\frac{|\tau|}{\beta}\right)^{\alpha}
  \xrightarrow{|\tau|\gg\beta}
  \sigma_{a}^{2}|\tau|^{\alpha},
    \label{eq:acc}
\end{equation}
with $\sigma_{a}^{2}\equiv \rho_{a}^{2}/\beta^{\alpha}$. Thus, maximizing the Shannon--Jaynes entropy under the two experimental constraints yields a probability density with a scale-free $\lambda$ tail and an exponential cutoff (Eq.~\eqref{eq:plambda}). The collective effect of the clutch traction pulses is to generate a power--law decay with a \emph{negative} exponent $\alpha<0$ in the acceleration autocorrelation function (Eq.~\eqref{eq:acc}). In this framework, the long-time decay is fixed by the small-$\lambda$ behavior of $p(\lambda)$ rather than by fitting the autocorrelation directly. The physics-informed Jeffreys prior $m(\lambda)$ does not assume a specific microscopic pulse shape. As shown in the Supplemental Material, the long-time exponent depends only on the small-$\lambda$ tail of $p(\lambda)$, while the detailed pulse shape affects only prefactors and short-time crossovers (see Secs.~S4 and~S5 of the Supplemental Material  \cite{SM}).

\section{Superdiffusive growth of axons on micropatterned substrates}\label{sec:Super}
Using the clutch mechanism and the MaxEnt principle we have derived an explicit model for axonal growth. Next, we show that this model accounts for the experimentally observed superdiffusive dynamics of axons on micropatterned PDMS substrates. We start with the Langevin Equation ~\eqref{eq:langevin} for growth cone position $x(t)$, where $a(t)$ is the stochastic acceleration with power‐law autocorrelation for large times $\tau\gg\beta$:
\begin{equation}
  \langle a(t)\,a(t+\tau)\rangle \;=\; \sigma_a^2\,\tau^\alpha,
  \label{eq:a_corr}
\end{equation}

To find the position $x(t)$, we integrate Equation ~\eqref{eq:langevin} twice to obtain the formal solution:
\begin{align}
  x(t) 
  &\;=\; \int_0^t \left[\int_0^u e^{-\gamma(u - s)}\,a(s)\,ds\right] \,du\nonumber\\
  &\;=\;
  \frac{1}{\gamma}
  \int_0^t \Bigl[\,1 - e^{-\gamma\,[\,t - s\,]}\Bigr]\,a(s)\,ds.
  \label{eq:x_t}
\end{align}
assuming zero initial velocity. 

\begin{figure*}[t!]
 \centering
\includegraphics[width=0.8\linewidth]{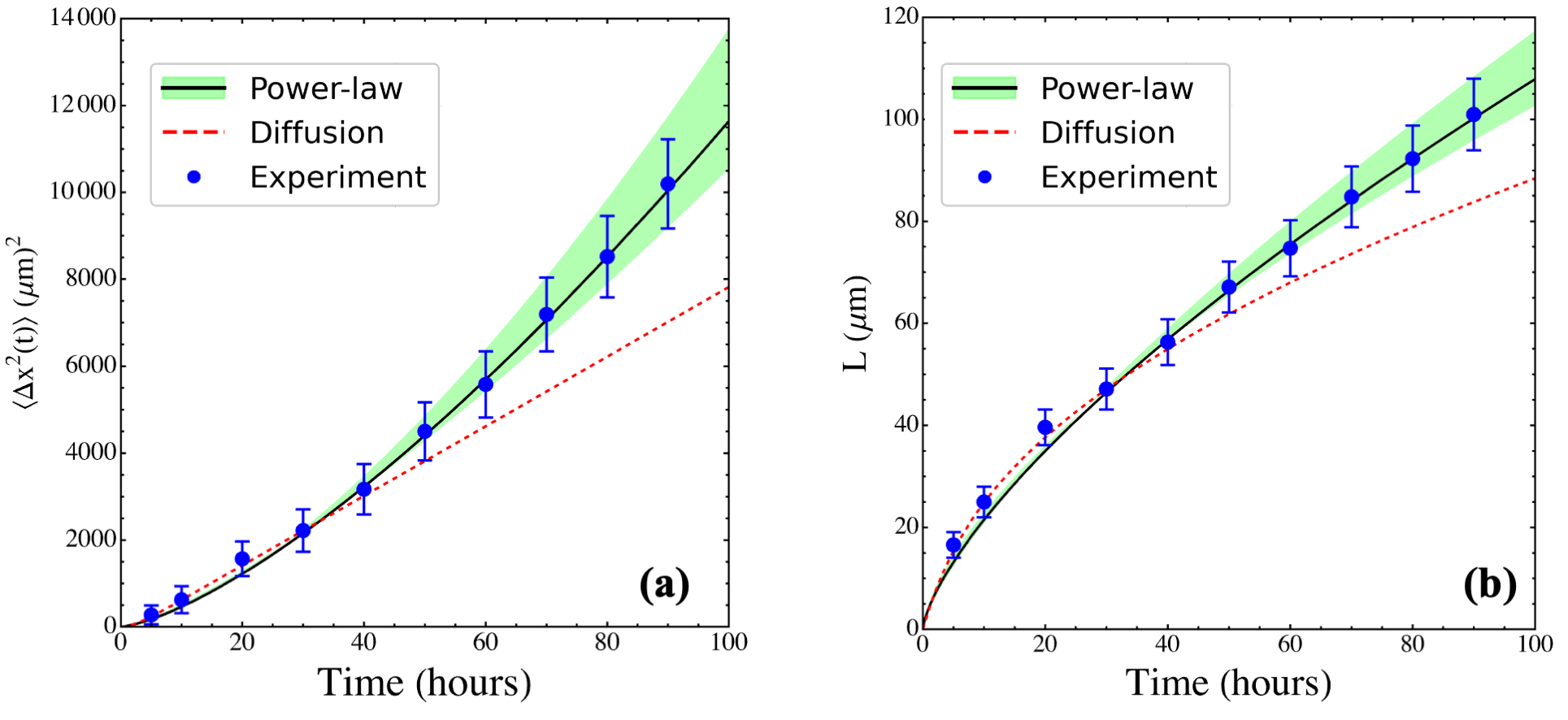}
\caption{
(a) Time evolution of the mean squared displacement (MSD) for axons growing on PDMS micropatterns with spatial period $d = 3~\mu\mathrm{m}$. Blue points show the MSD extracted from time-lapse fluorescence trajectories of $N=372$ axons. Error bars indicate the standard error of the mean. The red dotted line shows the pure diffusion prediction calculated using the measured diffusion coefficient $D = 20~\mu\mathrm{m}^{2}\,\mathrm{h}^{-1}$. The black solid line shows the prediction of the clutch-based stochastic model with exponent $\alpha=-0.6$, yielding $\langle x^{2}(t)\rangle \propto t^{2+\alpha}=t^{1.4}$. The narrow green band shows the model response for $-0.7 \le \alpha \le -0.5$, illustrating the sensitivity of the long-time scaling to variations in the power-law exponent.
(b) Time evolution of the mean axonal length $L$ for the same population of axons. Blue points show the experimental data, and error bars indicate the standard error of the mean. The red dotted line shows the corresponding pure diffusion prediction, while the black solid line shows the clutch-based stochastic model prediction for $\alpha=-0.6$. The narrow green band shows the model response for $-0.7 \le \alpha \le -0.5$.}
\label{fig4}
\end{figure*}

By using the power--law autocorrelation in Eq.~\eqref{eq:a_corr} and Eq.~\eqref{eq:x_t} we get (see Sec. S2 of the Supplemental Material \cite{SM}):
\begin{equation}
    \begin{split}
      &\langle x^2(t)\rangle
      \;=\;\\
      &\frac{\sigma_a^2}{\gamma^2}
      \int_0^t \!\int_0^t
        \Bigl[1 - e^{-\gamma \,[t - s]}\Bigr]
        \Bigl[1 - e^{-\gamma \,[t - s']}\Bigr]
        \,\bigl|\,s-s'\bigr|^\alpha
      \;ds\,ds'.
  \end{split}
  \label{eq:x2_general}
\end{equation}
This double integral completely specifies the mean‐-squared displacement (MSD) $\langle x^2(t)\rangle$ and the axonal mean squared length $L$ at any finite $t$. For large times $t\gg 1/\gamma$ we obtain (details in Sec.~S2 of the Supplemental Material \cite{SM}):
\begin{equation}
  \langle\Delta x^{2}(t)\rangle
  \simeq\frac{2\sigma_{a}^{2}}{\gamma^{2}}\,
        \frac{t^{2+\alpha}}{(1+\alpha)(2+\alpha)}.
        \label{eq:x2_asym}
\end{equation}

The power‐law growth, $\langle x^2(t)\rangle \propto t^{\alpha+2}$, indicates that the process can exhibit subdiffusion, superdiffusion, or normal diffusion, depending on the value of~$\alpha$. To determine this exponent, we build on the model of active biopolymers developed by Swanson and Wingreen, in which the joint probability distribution $p(x, L, t)$—for the cap size~$x$ and filament length~$L$—satisfies a Fokker–Planck equation with drift from polymerization and hydrolysis rates \cite{Swanson2011}. First-passage time (FPT) analysis yields a disassembly time distribution $P_{\mathrm{FPT}}(t) \propto t^{-3/2} \exp[-(at + 2)^2 / 4Dt]$, where the heavy-tailed $t^{-3/2}$ prefactor reflects broad filament relaxation times \cite{Swanson2011}.

During a fixed observation window the axon contains a renewing ensemble of
$N$ stochastically nucleated filaments. Linear response of the traction force
$F(t)$ to a single filament’s load step implies that:
\begin{equation}
  a(t)\;\propto\;F(t) = \sum_{i=1}^{N} f_i\,\mathrm e^{-\lambda_i t}\,\Theta(t-t_i)  
  \label{eq:a_traction}
\end{equation}
where each mode decays with a rate $\lambda_i\!\sim\!1/\text{lifetime}_i$. If we assume that the set
$\{\lambda_i\}$ in Eq.~\eqref{eq:a_traction} inherits its statistics from the FPT distribution, we can relate the probability density of rates via the mapping $p(\lambda)\,d\lambda = P_{\!\mathrm{FPT}}(t)\,dt$, with $t=1/\lambda$. This yields a rate distribution with a tail that behaves as $p(\lambda) \propto t^{1/2}$. When combined with the exact expression given by MaxEnt principle (Eq.~\eqref{eq:plambda}) this leads to a power--law exponent $\alpha=-1/2$. Therefore, the value of $\alpha$ is determined by the $t^{-3/2}$ tail of the disassembly time distribution characteristic for active actin and microtubule filaments.

For $\alpha=-1/2$, Eq.~\eqref{eq:x2_asym} predicts a MSD scaling as \(\langle x^2(t)\rangle \propto t^{1.5}\), which is in excellent agreement with the experimentally observed power-law behavior \(\langle x^2(t)\rangle \propto t^{1.4}\) measured using fluorescence imaging of axons growing on PDMS micropatterns \cite{Yurchenko2019}. This close match between the theoretical prediction and experiment provides strong support for the clutch-based stochastic model, in which transient adhesion events and the active dynamics of actin and microtubule filaments together govern axonal growth.

To link our theoretical framework to empirical data, we acquired time\textendash lapse fluorescence images of axons extending along PDMS micropatterned grooves with pattern spatial period $d = 3\,\mu\text{m}$ (Fig.~\ref{fig1}). Analysis of the ensemble of trajectories in the early\textendash time window ($t < 30\,\text{h}$) provides direct estimates of the diffusion (cell motility) coefficient and damping constant, yielding $D = 20\,\mu\text{m}^{2}\,\text{h}^{-1}$ and $\gamma = 0.1\,\text{h}^{-1}$, in agreement with our previous studies of axonal growth on similar substrates \cite{Descoteaux2022, Yurchenko2019, Basso2019, Vensi2019}. The parameter $\gamma$ captures the combined effects of cytoskeletal drag, membrane tension, and weak adhesion to the PDMS surface, while $D$ sets the magnitude of passive fluctuations. Together they define the diffusive baseline $\langle x^{2}(t)\rangle = 2Dt$ and the crossover time $t_{c} \sim \gamma^{-1}$ beyond which active processes dominate. Fig.~\ref{fig4}(a) contrasts this diffusion baseline (red dotted line) with the clutch\textendash based stochastic model (solid black line). The diffusive prediction matches the experimental MSD (blue dots) up to $t \approx 30\,\text{h}$; thereafter the data depart from the linear trend and follow the power\textendash law curve $\langle x^{2}(t)\rangle \propto t^{1.4}$ expected for $\alpha = -0.6$ (black solid line). The sharp crossover, captured without additional fitting, confirms that the measured $D$ and $\gamma$ establish the passive regime, whereas intermittent clutch engagements with heavy\textendash tailed lifetimes drive the observed super\textendash diffusive extension at longer times. Fig.~\ref{fig4}(b) shows the corresponding analysis for the mean axonal length $L$, which exhibits a similar crossover from the diffusion-dominated trend at early times to faster, clutch-mediated growth at longer times.

The narrow green bands surrounding the black power-law curves in
Fig.~\ref{fig4} illustrate the model's sensitivity to modest variations of the
correlation exponent in the range $-0.7 \le \alpha \le -0.5$. For example, the
two boundaries of the band in Fig.~\ref{fig4}(a) correspond to the MSD scaling
$\langle x^{2}(t)\rangle \propto t^{2+\alpha}$ evaluated at the limiting
exponents, while the central black line marks the nominal value
$\alpha=-0.6$. The experimental points remain within this shaded band for
$t>30~\mathrm{h}$, showing that small variations in $\alpha$ do not alter the
conclusion that the long-time growth lies well above the purely diffusive
baseline. This shaded region therefore provides a quantitative measure of the
robustness of the clutch-based prediction: even when the acceleration
correlation exponent deviates from its nominal value, the
superdiffusive regime persists and continues to capture the observed axonal
growth.

A statistical comparison of the post-crossover MSD data, $t\ge 30~\mathrm{h}$, with both a linear-in-time (diffusive) MSD baseline and the ballistic MSD scaling expected for simple drift dominated advance gives a fitted scaling exponent $\nu=1.395\pm0.056$. This exponent is significantly larger than the diffusive MSD exponent $\nu=1$ ($p=4.7\times10^{-6}$) and significantly smaller than the ballistic exponent $\nu=2$ expected for simple linear drift. Thus, the long-time MSD growth is not statistically consistent with either normal diffusion or simple linear drift, but instead supports a robust superdiffusive scaling regime (Sec.~S6 in the Supplemental Material \cite{SM}).

\begin{figure}
\centering
\includegraphics[width=1\linewidth]{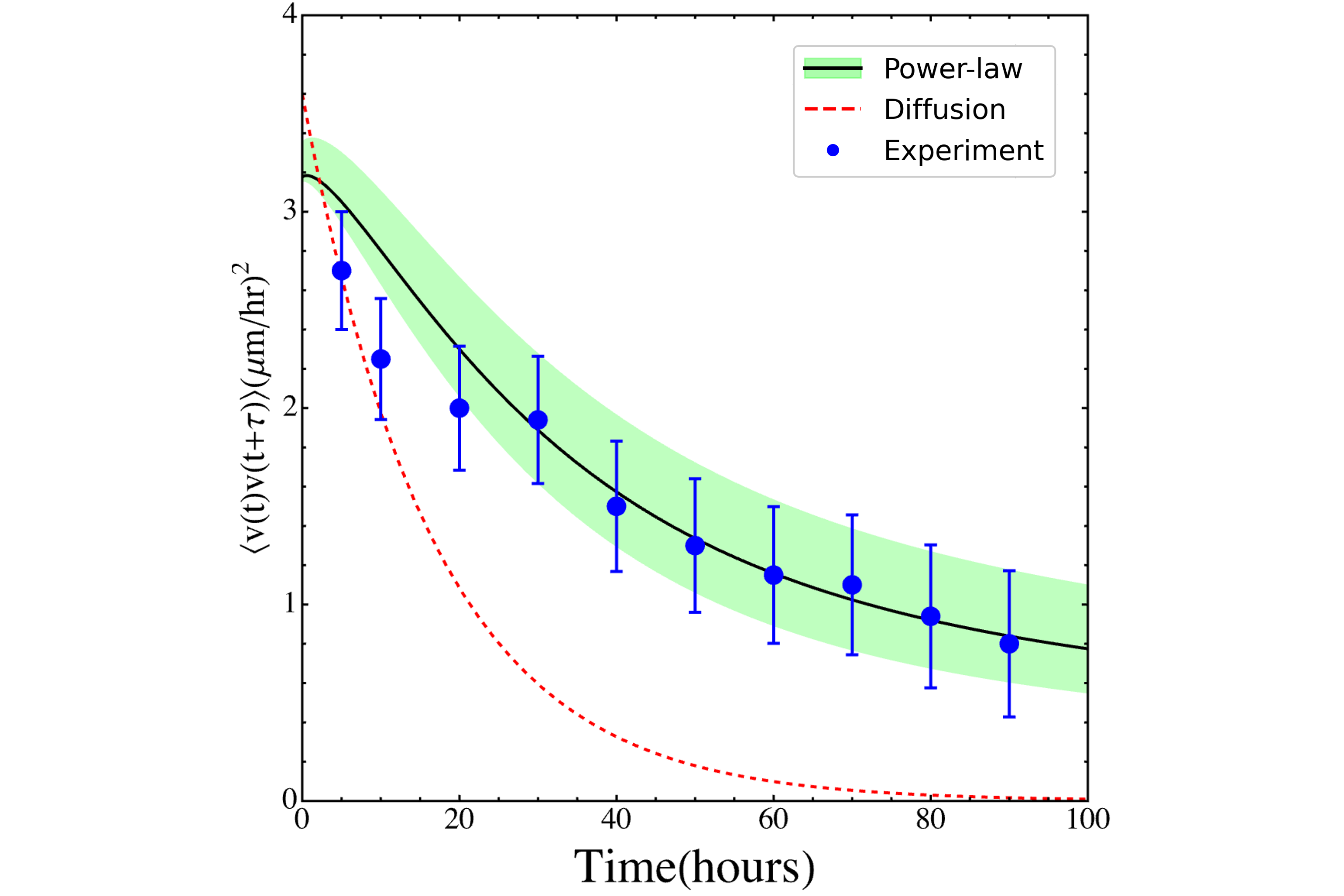}
\caption{Time dependence of the velocity autocorrelation function $C_v(\tau)$ for axonal growth on PDMS micropatterns with $d = 3\;\mu\text{m}$. Blue dots represent the experimental $C_v(\tau)$ obtained from the time-lapse imaging of the trajectories. Error bars indicate the standard error of the mean. Red dotted line: Ornstein–Uhlenbeck baseline $C_{v}(t) = 2D\gamma \,e^{-\gamma t}$ computed with measured $D = 20\;\mu\text{m}^{2}\,\text{h}^{-1}$ and $\gamma = 0.1\;\text{hr}^{-1}$.  Black solid line: clutch-based stochastic model with exponent $\alpha = -0.6$, giving the long-time power-law decay $C_{v}(t)\propto t^{-0.6}$. The narrow green band brackets the model response for $-0.7 \le \alpha \le -0.5$, showing the robustness of the predicted scaling against variations in the power--law exponent.}
\label{fig5}
\end{figure}

We next analyze the velocity–autocorrelation function, which characterizes the time‐dependent memory of the growth cone dynamics.  By integrating the Langevin Equation ~\eqref{eq:langevin} and using the power-law autocorrelation for acceleration (Eq.~\eqref{eq:a_corr}) we find that in the long–time limit ($\tau \gg \gamma^{-1}$)  
the velocity correlation function follows a power-law (explicit derivation is given in Sec.~S3 of the Supplemental Material \cite{SM}):
\begin{equation}
  C_v(\tau)\;\simeq\;\frac{\sigma_a^{2}}{\gamma^{2}}\,
  \tau^{\alpha},\qquad \alpha=-0.6,
  \label{eq:vv_asymptotic}
\end{equation}
showing that velocity fluctuations retain long‐range correlations even as their overall magnitude is attenuated by the square of the relaxation time $1/\gamma$. Eq.~\eqref{eq:vv_asymptotic} is the direct analogue of Eq.~\eqref{eq:x2_asym} for velocities: it couples the acceleration statistics to the exponential relaxation imposed by cytoskeletal damping and cell adhesion to the growth substrate. Fig.~\ref{fig5} compares this prediction (black curve, $\alpha=-0.6$) with the experimental $C_v(\tau)$ (blue dots) and with the purely diffusive OU baseline (red dotted line).  A narrow green band, calculated for $-0.5\le\alpha\le-0.7$, surrounds the theoretical curve, illustrating the robustness of the power–law decay: the data remain within this shaded region for $\tau>30$~h, confirming that small variations in adhesion statistics leave the long‐time behavior essentially unchanged. The results in Figs.~\ref{fig4} and ~\ref{fig5} concern ensemble-averaged observables across many axons on micropatterns. In particular, the observed power-law decay of the velocity autocorrelation function is fully consistent with single growth cone dynamics composed of near-constant velocity advance runs separated by pauses, as commonly seen in early axonal growth (see Secs.~S4 and ~S5 of the Supplemental Material \cite{SM}).

\section{Discussion} \label{sec:Discussion}

The results presented here establish an information-theoretic route for constructing an effective stochastic description of axonal growth fluctuations. Starting from a coarse-grained Langevin equation for growth cone motion, the colored acceleration noise is not specified phenomenologically. Instead, its effective relaxation-rate distribution is inferred from a constrained Shannon--Jaynes maximum entropy construction. The resulting probability density, Eq.~\eqref{eq:plambda}, generates a stationary colored-acceleration process whose autocorrelation is given by Eq.~\eqref{eq:a_corr}. This same noise spectrum determines the long-time scaling of both the mean squared displacement (Eq.~\eqref{eq:x2_asym}) and the velocity autocorrelation (Eq.~\eqref{eq:vv_asymptotic}). Thus, the model connects experimentally accessible constraints on growth cone dynamics to the anomalous transport laws measured for axons on micropatterned substrates.

A key point is that the MaxEnt inference is applied to the distribution of relaxation rates, rather than to the detailed time course of each microscopic traction event. Each relaxation mode contributes a kernel $e^{-\lambda|\tau|}$ to the acceleration autocorrelation, where $\lambda$ is a positive rate and $1/\lambda$ is the corresponding correlation time. These kernels preserve their form under a rescaling of time: if $\tau\mapsto c\tau$, the same kernel is recovered by rescaling the rate as $\lambda\mapsto \lambda/c$. Thus $\lambda$ acts as a positive scale parameter. The Jeffreys prescription then motivates a log-uniform baseline measure for $\lambda$, corresponding to equal weight per decade of relaxation rate. In the present setting, the observable acceleration autocorrelation is a mixture of such kernels, and the small-$\lambda$ sector is further constrained by the broad lifetime statistics of engaged traction intervals. This motivates the physics informed Jeffreys-type prior $m(\lambda)=\lambda^{-(1+\alpha)}$ used in the MaxEnt construction, which can be viewed as a log-uniform baseline modified by the small-$\lambda$ scaling associated with long-lived clutch or adhesion engagements. The prior therefore combines scale invariance for positive relaxation rates with independent biophysical information about long-lived traction events.

The long-time behavior follows directly from the small-$\lambda$ part of the inferred distribution. For the molecular clutch model introduced in \hyperref[sec:clutch]{Section~III}, the acceleration autocorrelation is proportional to the Laplace transform of the relaxation rate distribution, as expressed in Eq.~\eqref{eq:Ctau}. Substituting the MaxEnt distribution, Eq.~\eqref{eq:plambda}, gives the power-law autocorrelation in Eq.~\eqref{eq:acc}. The exponent must satisfy $\alpha<0$ so that the rate distribution is normalizable at small $\lambda$, since $p(\lambda)\propto \lambda^{-(1+\alpha)}$ in this limit. Consequently, the acceleration autocorrelation decays as a power law at long times,
and the MSD scaling becomes $\langle \Delta x^2(t)\rangle\propto t^{2+\alpha}$, as shown in Eq.~\eqref{eq:x2_asym}. This derivation makes explicit why the anomalous exponent is controlled by the low rate, long-lifetime part of the relaxation spectrum rather than by the detailed shape of individual traction pulses.

The exponent is also robust to a broad class of microscopic and statistical variations. If the prior is multiplied by a smooth factor, $m(\lambda)\mapsto m(\lambda)q(\lambda)$, where $q(\lambda)$ remains finite and nonzero as $\lambda\to0^+$, the MaxEnt solution retains the same small-$\lambda$ power law. Likewise, replacing a single-exponential relaxation kernel by a finite-duration pulse, such as a ramp-and-terminate traction event, changes only prefactors and short-time crossovers  (Supplemental Material, Secs.~S4 and ~S5 \cite{SM}). Thus, the asymptotic power law remains fixed by the small-$\lambda$ behavior of $p(\lambda)$. This robustness is important physically because clutch-mediated traction is unlikely to be described exactly by a single microscopic pulse shape. What matters for the long-time transport is the distribution of relaxation or engagement times, not the detailed temporal profile of any one event.

The biophysical interpretation of the exponent is therefore naturally expressed in terms of lifetime statistics. Adhesion turnover, clutch engagement and disengagement, actomyosin contractility, and cytoskeletal remodeling all contribute to intermittent force transmission at the growth cone \cite{Lowery2009, Franze2010, Franze2020, Huber2003}. Broadly distributed engagement times map, through $\lambda=1/T$, to a small-$\lambda$ tail in the relaxation-rate distribution. One microscopic anchoring for such a tail is provided by active biopolymer first-passage statistics, which motivates the representative value of the power law exponent $\alpha\simeq -1/2$ \cite{Swanson2011}. Figs.~\ref{fig4} and~\ref{fig5} demonstrate that the experimental data are well described by the nearby value $\alpha\simeq -0.6$. This agreement does not by itself identify a unique molecular pathway. Rather, it indicates that the measured anomalous transport is consistent with a class of microscopic mechanisms that generate heavy-tailed engagement or relaxation times.

This coarse-grained description clarifies how distinct microscopic descriptions can give rise to the same measured ensemble observables. A mixture of OU-like relaxation modes, a collection of finite duration traction pulses, or an advance--pause description with broadly distributed dwell times can all yield the same long time exponent if they share the same small-$\lambda$ behavior. The measured scaling exponent therefore identifies an effective dynamical class rather than a unique single-axon mechanism. The MaxEnt construction provides a compact representation of this class while keeping the inference tied to measurable constraints. In this sense, the theory complements detailed mechanochemical models: it does not replace molecular descriptions of the clutch, cytoskeleton, or adhesion complexes, but provides the statistical link needed to connect those processes to long time growth laws.

The comparison with experiment supports this analysis. At early times, the measured MSD follows the diffusive baseline determined by the independently estimated diffusion coefficient and damping constant. At later times, the data depart from the linear diffusive trend and follow the power law behavior predicted by the colored acceleration model, as shown in Fig.~\ref{fig4}(a). The corresponding mean axonal length exhibits the same crossover from diffusion dominated behavior to faster clutch mediated extension (Fig.~\ref{fig4}(b)). The velocity autocorrelation in Fig.~\ref{fig5} provides an independent test because it probes temporal memory directly rather than only the time-integrated displacement. Its power law decay is consistent with the exponent that controls the MSD, showing that the anomalous growth is associated with persistent correlations in the effective traction noise. The shaded bands in Figs.~\ref{fig4} and~\ref{fig5} further show that small variations in $\alpha$ preserve the long-time behavior while remaining clearly distinct from the purely diffusive OU baseline.

The analytical form of the model suggests several experimental tests. Perturbations that alter adhesion turnover, actomyosin activity, ATP availability, or microtubule stability should modify the inferred relaxation-rate distribution. In the Langevin description, such perturbations may change the noise amplitude, shift the crossover time, alter the effective damping, or modify the exponent $\alpha$. For example, inhibition of myosin-II activity using blebbistatin, perturbation of microtubule dynamics using taxol or related compounds, or direct modulation of adhesion turnover should affect both $\langle \Delta x^2(t)\rangle$ and $C_v(\tau)$ \cite{Basso2019, Descoteaux2022, Kumarasinghe2022}. A useful test would be to determine whether the exponent inferred from the MSD changes consistently with the exponent inferred from the velocity autocorrelation under the same perturbation. Such a comparison would help distinguish changes in fluctuation amplitude from changes in the underlying distribution of engagement times.

The results also have broader implications for biological physics. In many active biological systems, stochastic motion is driven by internal energy consumption, molecular turnover, and intermittent force generation. The noise in such systems is not simply a passive environmental background, but reflects the active processes that drive motion. Treating this noise as a quantity to be inferred rather than as an externally prescribed input provides a way to formulate effective theories when long-time trajectories are measurable but microscopic dynamics are only partially resolved. Similar issues arise in cell migration, collective cellular motion, tissue remodeling, intracellular transport, and morphogenesis \cite{Dieterich2008, Alert2020, Oliveri2022, amselem2012stochastic, Metzler2000, Murray1993, Edwards2007}. The MaxEnt-based approach developed here may therefore be useful for connecting experimentally tunable variables, such as substrate stiffness, topography, adhesion chemistry, or cytoskeletal activity, to emergent fluctuation statistics and coarse-grained dynamical laws.

The same framework also links axonal growth to active matter systems with temporally correlated fluctuations. Axonal growth on micropatterned substrates provides a biological example of anomalous transport generated by internally driven, temporally correlated fluctuations. Related power law statistics and long-memory effects occur in active colloids, persistent random walks, L\'evy-like search processes, and other nonequilibrium systems with broadly distributed relaxation or persistence times \cite{Peruani2007, Tilles2017, Reynolds2012, Metzler2000}. The present analysis shows how such scaling can arise from an entropy-constrained spectrum of active force fluctuations. It therefore places axonal growth within a wider class of driven systems in which macroscopic transport is governed primarily by the distribution of event durations and relaxation rates, rather than by the detailed form of individual microscopic events.

The model is intrinsically coarse grained. It focuses on ensemble-level statistics of the longest neurite on aligned micropatterns and does not describe neurite branching, axon specification, biochemical guidance gradients, or cell--cell interactions. It also assumes that the effective acceleration noise is stationary over the observation window and that the long-time behavior is controlled by the small-$\lambda$ tail of the inferred distribution. These assumptions are appropriate for the measurements analyzed here, but they may require modification for dense cultures, three-dimensional matrices, strongly time-dependent environments, or substrates with more complex geometries. Future extensions could incorporate anisotropic two-dimensional motion, substrate-dependent relaxation spectra, coupling between axonal length and traction generation, or nonstationary MaxEnt constraints. Such extensions would test how broadly entropy-constrained noise can describe growth fluctuations in more complex neuronal environments.

\section{Conclusions}\label{sec:Conclusions}

We have developed a coarse grained stochastic theory for axonal growth on micropatterned substrates using the Shannon--Jaynes maximum entropy principle. Starting from a Langevin description of growth cone motion, we inferred the effective distribution of traction-force relaxation rates from experimentally motivated constraints rather than postulating the colored noise directly. The resulting colored acceleration process exhibits power law temporal correlations and yields closed form predictions for both the mean squared displacement and the velocity autocorrelation. The theory accounts for the observed crossover from early diffusive behavior to long-time superdiffusive growth in cortical neurons on micropatterned PDMS substrates, including the measured mean squared displacement scaling with exponent $1.4$ and the corresponding power law decay of the velocity autocorrelation.

The physical origin of the anomalous scaling is the broad distribution of effective relaxation or engagement times associated with clutch mediated traction and cytoskeletal remodeling. In this framework, the superdiffusive exponent emerges from the inferred small-$\lambda$ behavior of the relaxation rate distribution, rather than being imposed as a fitting parameter. More broadly, the results illustrate how information-theoretic inference can connect partially observed microscopic activity to macroscopic growth laws in biological systems. Future perturbation experiments targeting adhesion turnover, actomyosin contractility, ATP availability, or microtubule dynamics can test whether changes in the inferred relaxation spectrum accompany changes in the MSD, axonal length, and velocity autocorrelation. Thus, entropy-constrained stochastic modeling provides a useful route for linking molecular perturbations to emergent growth behavior in neuronal systems and active matter.

\vspace{0.5\baselineskip}

\begin{acknowledgments}
The authors acknowledge financial support for this work from
National Science Foundation (DMR 2104294), Tufts Faculty Award (C.S.), and Tufts Summer Scholars Program (J.S.).  
\end{acknowledgments}

\bibliography{references}

\clearpage
\onecolumngrid

\begin{center}
{\bf\large{Supplemental Material for Maximum-Entropy Model of Colored Noise in Superdiffusive Axonal Growth}}
\end{center}
\setcounter{section}{0}
\setcounter{subsection}{0}
\setcounter{equation}{0}
\renewcommand{\theequation}{S\arabic{equation}}
\setcounter{secnumdepth}{2} 
\renewcommand{\thesection}{S\arabic{section}}
\textbf{This PDF file includes:}

S1. Solution to the variational problem

S2. Derivation of the power-law scaling of the mean squared displacement

S3. Analysis of the velocity autocorrelation function

S4. Proof that the long-time correlation exponent is independent of pulse build and cutoff

S5. From ballistic-pausing motion of single axons to superdiffusive ensemble dynamics

S6. Statistical comparison with diffusive and ballistic MSD scaling

\section{Solution of the variational problem}

The goal is to determine the rate distribution \(p(\lambda)\) (\(\lambda>0\))
that maximizes the Shannon--Jaynes entropy
\begin{equation}
  S[p]=
  -\int_{0}^{\infty} d\lambda\;
    p(\lambda)\,\ln\!\frac{p(\lambda)}{m(\lambda)},
  \qquad
  m(\lambda)=\lambda^{-(1+\alpha)}.
\end{equation}

subject to the constraints

\begin{align}
  &\text{normalisation:}      &
  \int_{0}^{\infty} p(\lambda)\,d\lambda &= 1, \label{eq:norm}\\
  &\text{finite mean relaxation rate:} &
  \int_{0}^{\infty}\lambda\,p(\lambda)\,d\lambda &= \langle\lambda\rangle, 
                                                             \label{eq:mean}
\end{align}

Introducing Lagrange multipliers \(\chi\) and \(\beta\) for these two constraints, and the Jeffreys prior \(m(\lambda)=\lambda^{-(1+\alpha)}\) (log-uniform baseline modified by the small-$\lambda$ scaling implied by the heavy-tailed lifetimes of active biopolymers), we extremize the functional:

\begin{equation}
  \mathcal{J}[p]=S[p]
       -\chi\Bigl(\int p(\lambda)\,d\lambda-1\Bigr)
       -\beta\Bigl(\int \lambda p(\lambda)\,d\lambda-\langle\lambda\rangle\Bigr).
\end{equation}

Varying \(p\to p+\delta p\), with arbitrary \(\delta p(\lambda)\), and using 
\[
\delta S = -\int_{0}^{\infty} d\lambda\,\delta p(\lambda)
\left[\ln\frac{p(\lambda)}{m(\lambda)}+1\right],
\]
we obtain
\[
  \delta \mathcal{J}
  =\int_{0}^{\infty} d\lambda\;
     \delta p(\lambda)\,
     \left[
        -\ln\frac{p(\lambda)}{m(\lambda)}-1
        -\chi-\beta\lambda
     \right].
\]
Because the bracket must vanish for arbitrary \(\delta p(\lambda)\) we get
\[
  \ln\frac{p(\lambda)}{m(\lambda)}
    = -1-\chi-\beta\lambda.
\]
Hence
\[
  p(\lambda)=e^{-1-\chi}m(\lambda)e^{-\beta\lambda}.
\]
Substituting the prior \(m(\lambda)=\lambda^{-(1+\alpha)}\) gives
\begin{equation}
  p(\lambda)
  =A\,\lambda^{-(1+\alpha)}e^{-\beta\lambda},
  \qquad
  A=e^{-1-\chi}.
\end{equation}
Using the normalization condition, the constant \(A\) can be written equivalently as \(A=1/Z\), so that:
\begin{equation}
  p(\lambda)
  =\frac{1}{Z}\lambda^{-(1+\alpha)}e^{-\beta\lambda},
  \qquad
  Z=\int_{0}^{\infty}\lambda^{-(1+\alpha)}e^{-\beta\lambda}\,d\lambda.
  \label{eq:solution}
\end{equation}
Next, we show that the normalization integral $Z$ in Equation ~\eqref{eq:solution} reduces to the Gamma-like form:
\[
  Z=\beta^{\alpha}\,\Gamma(-\alpha).
\]

The normalisation integral is:
\[
  Z
  =\int_{0}^{\infty}\lambda^{-(1+\alpha)}e^{-\beta\lambda}\,d\lambda.
\]
The integral converges both at \(\lambda\to0^{+}\) and \(\lambda\to\infty\) if $\beta>0$ and \(\alpha<0\). Let \(\rho\equiv-\alpha>0\), then
\[
  Z
  =\int_{0}^{\infty}\lambda^{\rho-1}e^{-\beta\lambda}\,d\lambda.
\]
Introducing the dimensionless variable \(u=\beta\lambda\):
\[
  \lambda=\frac{u}{\beta},
  \qquad
  d\lambda=\frac{du}{\beta}.
\]
Hence
\[
  Z
  =\int_{0}^{\infty}
       \bigl(\tfrac{u}{\beta}\bigr)^{\rho-1}
       e^{-u}\,
       \frac{du}{\beta}
  =\beta^{-\rho}\int_{0}^{\infty}u^{\rho-1}e^{-u}\,du.
\]

By definition of the Gamma function,
\(\displaystyle\Gamma(\rho)=\int_{0}^{\infty}u^{\rho-1}e^{-u}\,du\)
for \(\rho>0\).  Therefore
\begin{equation}
    Z
  =\beta^{-\rho}\Gamma(\rho)
  =\beta^{\alpha}\Gamma(-\alpha),
  \label{eq:GammaF}
\end{equation}
re-expressing \(\rho=-\alpha\) in the final step.  
The conditions \(\alpha<0\) and \(\beta>0\) guarantee convergence at both
integration limits.

The fixed mean \(\langle\lambda\rangle\) determines the scale \(\beta\):
\[
  \langle\lambda\rangle
  =\frac{1}{Z}\int_{0}^{\infty}
      \lambda^{-\alpha}\,e^{-\beta\lambda}\,d\lambda
  =\frac{\Gamma(1-\alpha)}{\beta\,\Gamma(-\alpha)}
  \;\Longrightarrow\;
  \beta=\frac{\Gamma(1-\alpha)}{\langle\lambda\rangle\,\Gamma(-\alpha)}.
\]

Note that all Lagrange multipliers are now fixed; no further structure can be
added without violating the maximal-entropy condition.


\section{Derivation of the power law scaling of the mean squared displacement}

We start with  the stochastic Langevin equation:
\begin{equation}
  \ddot{x}(t) \;=\; -\gamma\,\dot{x}(t) \;+\; a(t),
  \label{eq:eom}
\end{equation}
where $a(t)$ is a random acceleration with power‐law autocorrelation:
\begin{equation}
  \langle a(t)\,a(t+\tau)\rangle \;=\; \sigma_a^2\,\tau^\alpha,
  \label{eq:a_corr}
\end{equation}
for $\tau>0$, and $\gamma>0$ is a damping constant.  We wish to find the mean‐square displacement $\langle x^2(t)\rangle$.

\paragraph{Solution for $x(t)$.} First, rewrite \eqref{eq:eom} for the velocity $v(t) := \dot{x}(t)$:
\[
  \dot{v}(t) + \gamma\,v(t) \;=\; a(t),
\]
Integrate to get the formal solution
\[
  v(t) \;=\; \int_0^t e^{-\gamma\,[\,t - s\,]}\,a(s)\,ds,
\]
assuming zero initial velocity.  Then
\[
  x(t) \;=\; \int_0^t v(u)\,du
  \;=\; \int_0^t \left[\int_0^u e^{-\gamma(u - s)}\,a(s)\,ds\right] \,du.
\]
One can interchange integrals to obtain:
\[
  x(t)
  \;=\;
  \frac{1}{\gamma}
  \int_0^t \Bigl[\,1 - e^{-\gamma\,[\,t - s\,]}\Bigr]\,a(s)\,ds.
\]

\paragraph{Expression for $\langle x^2(t)\rangle$}

Since $x(t)$ is a linear functional of $a(s)$, we have
\[
  x(t)
  \;=\;
  \int_0^t f(t-s)\,a(s)\,ds,
  \quad
  \text{where}
  \quad
  f(\tau) \;=\;\frac{1}{\gamma}\bigl[\,1 - e^{-\gamma\,\tau}\bigr].
\]
Then the mean‐square displacement is
\begin{align*}
  \langle x^2(t)\rangle
  &=
  \Bigl\langle
    \int_0^t f(t-s)\,a(s)\,ds
    \;\times\;
    \int_0^t f(t-s')\,a(s')\,ds'
  \Bigr\rangle
  \\[6pt]
  &= \int_0^t \!\int_0^t f(t-s)\,f(t-s')\;\langle a(s)\,a(s')\rangle\;ds\,ds'.
\end{align*}
Using the given correlation
\[
  \langle a(s)\,a(s')\rangle \;=\; \sigma_a^2\,|s-s'|^{\,\alpha},
\]
 we get
\begin{equation}
  \langle x^2(t)\rangle
  \;=\;
  \sigma_a^2
  \int_0^t \!\int_0^t
    f(t-s)\,f(t-s')\;\bigl|s-s'\bigr|^\alpha
  \;ds\,ds'.
  \label{eq:x2_general}
\end{equation}
Substituting $f(\tau)=\frac{1}{\gamma}\bigl[1 - e^{-\gamma \tau}\bigr]$, we find
\[
  \langle x^2(t)\rangle
  \;=\;
  \frac{\sigma_a^2}{\gamma^2}
  \int_0^t \!\int_0^t
    \Bigl[1 - e^{-\gamma \,[t - s]}\Bigr]
    \Bigl[1 - e^{-\gamma \,[t - s']}\Bigr]
    \,\bigl|\,s-s'\bigr|^\alpha
  \;ds\,ds'.
\]
This double integral completely specifies the mean‐square displacement at any finite $t$.

\textit{Large \it {t} asymptotics:}

For times $t$ large compared to $1/\gamma$, the exponentials $e^{-\gamma\,[t-s]}$ become negligible unless $s$ is very close to $t$.  Hence, for $t\gg 1/\gamma$, we can approximate
\[
  1 - e^{-\gamma\,[\,t - s\,]}
  \;\approx\; 1
  \quad
  (\text{for most }0 \le s \ll t).
\]
Thus,
\[
  \langle x^2(t)\rangle
  \;\approx\;
  \frac{\sigma_a^2}{\gamma^2}
  \int_0^t\!\!\int_0^t
    \bigl|s-s'\bigr|^\alpha\,
  ds\,ds'.
\]
and the double integral becomes:
\[
  \int_0^t\!\!\int_0^t |s-s'|^\alpha\,ds\,ds'
  \;=\;
  2 \int_0^t ds
     \int_0^s (s - s')^\alpha \,ds'
  \;=\;
  2 \int_0^t \frac{s^{\alpha+1}}{\alpha+1}\,ds
  \;=\;
  2\,\frac{t^{\alpha+2}}{(\alpha+1)\,(\alpha+2)}.
\]
Hence, for $t\gg 1/\gamma$,
\[
  \langle x^2(t)\rangle
  \;\sim\;
  \frac{2\,\sigma_a^2}{\gamma^2}\,\frac{t^{\,\alpha + 2}}{(\alpha+1)\,(\alpha+2)}.
\]
This power‐law growth, $\langle x^2(t)\rangle \propto t^{\alpha+2}$, shows that the process may exhibit sub-, super-, or normal diffusion (and even super-ballistic) depending on the value of~$\alpha$.

\subsection*{Summary}

\noindent
\textbf{Exact integral form:}
\[
  \langle x^2(t)\rangle
  \;=\;
  \frac{\sigma_a^2}{\gamma^2}
  \int_0^t \int_0^t
    \bigl[1 - e^{-\gamma (\,t-s\,)}\bigr]
    \bigl[1 - e^{-\gamma (\,t-s'\,)}\bigr]
    \,\lvert s - s' \rvert^\alpha
  \;ds\,ds'.
\]

\noindent
\textbf{Asymptotics for large $t$:}
\[
  \langle x^2(t)\rangle
  \;\sim\;
  \frac{2\,\sigma_a^2}{\gamma^2\,(\alpha+1)\,(\alpha+2)}
  \;t^{\alpha+2}.
\]

\section{Analysis of the velocity autocorrelation function}

From Eq.\ \eqref{eq:eom}, 
\[
  v(t) \;=\; \dot{x}(t) 
  \;=\; \int_{0}^{t} e^{-\gamma\,[\,t-s\,]}\,a(s)\,ds,
\]
assuming \(v(0)=0\).  We want to calculate 
\(\langle v(t)\,v(t+\tau)\rangle\).  Thus,
\begin{align*}
  \langle v(t)\,v(t+\tau)\rangle
  \;&=\;
  \Bigl\langle
    \int_{0}^{t} e^{-\gamma\,(t-s)}\,a(s)\,ds
    \;\times\;
    \int_{0}^{t+\tau} e^{-\gamma\,\bigl[\,(t+\tau)-s'\bigr]}\,a(s')\,ds'
  \Bigr\rangle
  \\[6pt]
  &= \int_{0}^{t}\!\!\int_{0}^{t+\tau}
    e^{-\gamma\,(t-s)}\,e^{-\gamma\,\bigl[\,(t+\tau)-s'\bigr]}
    \,\langle a(s)\,a(s')\rangle 
  \;ds\,ds'.
\end{align*}
Using \(\langle a(s)\,a(s')\rangle = \sigma_a^{2}\,|s-s'|^{\,\alpha}\), we get
\begin{equation}
  \langle v(t)\,v(t+\tau)\rangle
  \;=\;
  \sigma_a^{2}\,
  \int_{0}^{t}\!\!\int_{0}^{t+\tau}
  e^{-\gamma\,(t-s)}\,e^{-\gamma\,\bigl[\,(t+\tau)-s'\bigr]}
  \,\bigl|\,s - s'\bigr|^{\alpha}
  \;ds\,ds'.
  \label{eq:vv_correlation}
\end{equation}
This is the velocity autocorrelation for the finite‐time (non‐stationary, i.e. axons growing) case.

\paragraph{Long $t$ limit:}

In this limit, define the stationary autocorrelation function
\[
  C_{v}(\tau)
  \;=\;
  \lim_{\tau\to\infty}
  \langle v(t)\,v(t+\tau)\rangle.
\]
The resulting integral typically depends on \(\alpha\) in a nontrivial way, and closed‐form solutions are only expressible in terms of special functions. 

\textit{Asymptotic Evaluation of the Velocity Autocorrelation Function $C_v(\tau)$}

In Equation \eqref{eq:vv_correlation} define:
\begin{equation}
x=t-s, \qquad y=t+\tau - s',
\end{equation}
Equation (\ref{eq:vv_correlation}) becomes:
\begin{equation}
C_v^{(t)}(\tau)=\sigma_a^{2} \int_{0}^{t} dx \int_{0}^{t+\tau} dy \, e^{-\gamma(x+y)} \, \bigl | {\tau + x - y}\bigr |^{\alpha}.
\label{eq:Cv_xy}
\end{equation}

For $\tau \gg \gamma^{-1}$ we have:

\begin{equation}
C_v(\tau) \simeq \sigma_a^{2}\,\tau^{\alpha}
\left[\int_{0}^{t} dx\, e^{-\gamma x}\right]
\left[\int_{0}^{t+\tau} dy\, e^{-\gamma y}\right].
\label{eq:Cv_fac}
\end{equation}

The integrals in Eq.(\ref{eq:Cv_fac}) evaluate to:
\begin{equation}
\int_{0}^{t} dx\, e^{-\gamma x}=\frac{1-e^{-\gamma t}}{\gamma}, \qquad
\int_{0}^{t+\tau} dy\, e^{-\gamma y}=\frac{1-e^{-\gamma(t+\tau)}}{\gamma}.
\end{equation}
and substitution back into Eq.(\ref{eq:Cv_fac}) gives:
\begin{equation}
C_v(\tau) = \sigma_a^{2}\,\tau^{\alpha}
\frac{\bigl[1-e^{-\gamma t}\bigr]\bigl[1-e^{-\gamma(t+\tau)}\bigr]}{\gamma^{2}}.
\label{eq:Cv_prelimit}
\end{equation}
which in the limit $(t\gg\gamma^{-1})$ reduces to:
\begin{equation}
C_v(\tau) \simeq
\frac{\sigma_a^{2}}{\gamma^{2}}\, \tau^{\alpha}.
\label{eq:Cv_asymp}
\end{equation}
as claimed in Equation (14) in the main text.

\section{Proof that the long-time correlation exponent is independent of pulse build and cutoff}

We show in detail, that the long-time decay of the acceleration autocorrelation
$C_a(\tau)$, and therefore the population MSD exponent, is determined solely by the small–$\lambda$ behavior of the rate spectrum $p(\lambda)$, not by the micro-shape of a single traction pulse during clutch engagement. In particular, pulses that build during
engagement and end abruptly (ramp-and-terminate pulses) yield the same long-time exponent as the OU-like
exponential kernel when mixed over the same $p(\lambda)$.

\medskip
 Let each clutch engagement event generate a displacement pulse
$g_\lambda(t)$ over its lifetime, with an effective rate (inverse lifetime)
$\lambda>0$. The acceleration kernel is the time derivative,
$a_\lambda(t)=\dot g_\lambda(t)$, and we model the observed acceleration as a stationary
superposition of such pulses with independent random onsets and a rate spectrum
$p(\lambda)$.
For such a shot-noise process, the acceleration
autocorrelation can be written as a superposition over rates \cite{Murray1993}:
\begin{equation}
C_a(\tau)\;=\;\int_{0}^{\infty} p(\lambda)\,\kappa(\lambda)\,\phi_\lambda(\tau)\,d\lambda,
\qquad \phi_\lambda(\tau)\equiv
\frac{\int_{-\infty}^{\infty} a_\lambda(u)\,a_\lambda(u+\tau)\,du}
     {\int_{-\infty}^{\infty} a_\lambda^2(u)\,du},
\label{eq:S17}
\end{equation}
where $\kappa(\lambda)$ absorbs the (rate-dependent) pulse energy and event rate. We now compute $\phi_\lambda(\tau)$ for two physically relevant pulse shapes and show that, for
large $|\tau|$, both reduce to the same exponential dependence $\propto e^{-c\,\lambda|\tau|}$
(up to bounded prefactors). This suffices to fix the long-time scaling of $C_a(\tau)$ once
$p(\lambda)$ is specified.

\medskip
\noindent\textbf{Example A: OU-like pulse (reference).}
Take a single-mode OU acceleration kernel
$a_\lambda(t)=\lambda\,e^{-\lambda t}\,\Theta(t)$ (causal decay).
Then
\[
\int_{-\infty}^{\infty} a_\lambda(u)\,a_\lambda(u+\tau)\,du
=\lambda^2\int_{0}^{\infty} e^{-\lambda u}\,e^{-\lambda(u+|\tau|)}\,du
=\frac{\lambda}{2}\,e^{-\lambda|\tau|},
\]
and $\int a_\lambda^2=\lambda/2$. Hence
\begin{equation}
\phi_\lambda^{\text{(OU)}}(\tau)=e^{-\lambda|\tau|}.
\label{eq:phiOU}
\end{equation}

\medskip
\noindent\textbf{Example B: Amplitude build then abrupt-off (ramp-and-terminate) pulse}.
Let the displacement during a single engagement (lifetime $T$) build as
\[
g_{\lambda,T}(t)=(1-e^{-\lambda t})\,\Theta(t)\,\Theta(T-t),
\]
and vanish after detachment ($t\ge T$). The corresponding acceleration kernel is
\[
a_{\lambda,T}(t)=\dot g_{\lambda,T}(t)
=\lambda e^{-\lambda t}\,\Theta(t)\,\Theta(T-t)\,,
\]
i.e.\ an exponential increase during engagement and exactly zero after detachment (abrupt end). 
 Its autocorrelation is:
\[
\int_{-\infty}^{\infty} a_{\lambda,T}(u)\,a_{\lambda,T}(u+\tau)\,du
=\lambda^2\!\!\!\int_{0}^{\max(0,T-\tau)}\!\!\! e^{-\lambda u}\,e^{-\lambda(u+\tau)}\,du
=\frac{\lambda}{2}\,e^{-\lambda\tau}\Big(1-e^{-2\lambda(T-\tau)}\Big)_+,
\]
where the symbol $+$ is defined as: $(x)_+=\max\{x,0\}$.
The pulse integrated squared amplitude is $\int a_{\lambda,T}^2=\frac{\lambda}{2}(1-e^{-2\lambda T})$.
Therefore the normalized single-pulse autocorrelation is
\begin{equation}
\phi_{\lambda,T}^{\text{(build+cut)}}(\tau)
=\frac{e^{-\lambda|\tau|}\,\big(1-e^{-2\lambda(T-|\tau|)}\big)_+}
      {1-e^{-2\lambda T}}
=\begin{cases}
e^{-\lambda|\tau|}\,\dfrac{1-e^{-2\lambda(T-|\tau|)}}{1-e^{-2\lambda T}},
& 0\le |\tau|< T,\\[6pt]
0, & |\tau|\ge T.
\end{cases}
\label{eq:phicut}
\end{equation}
\textbf{Notes}:
(i) Strictly, $a_{\lambda,T}(t)=\lambda e^{-\lambda t}\Theta(t)\Theta(T-t)-\big(1-e^{-\lambda T}\big)\delta(t-T)$; the Dirac $\delta$ term contributes only contact terms at $\tau\simeq 0$ (or $\pm T$) in $\int a(u)a(u+\tau)\,du$ and does not alter the large $|\tau|$ decay, so omitting it leaves $C_a(\tau)$ and the resulting MSD exponent unchanged.
(ii) For time lags $|\tau|$ that are large on the rise timescale ($|\tau|\gg \lambda^{-1}$)
but still smaller than the lifetime ($|\tau|<T$), one has
$\phi_{\lambda,T}^{\text{(build+cut)}}(\tau)=e^{-\lambda|\tau|}\,[1+\mathcal O(e^{-2\lambda(T-|\tau|)})]$.
(iii) The abrupt cutoff at $|\tau|=T$ simply truncates correlations beyond the engagement
lifetime; when we mix over many events with a spectrum of lifetimes/rates, the small–$\lambda$
sector (long $T$) controls the large $|\tau|$ tail of $C_a(\tau)$.

Combining \eqref{eq:phiOU} and \eqref{eq:phicut}, we have the generic long-time form:
\begin{equation}
\phi_\lambda(\tau)\sim A(\lambda)\,e^{-c\,\lambda|\tau|}\,
\label{eq:phigen}
\end{equation}
with bounded $A(\lambda)$ and $c\approx1$.

\medskip
\noindent\textbf{Derivation of the long-time decay of $C_a(\tau)$.}
Inserting \eqref{eq:phigen} into \eqref{eq:S17}, we note that for large $|\tau|$ the integral is controlled by
small rates $\lambda$ (long-lived engagements). For heavy-tailed engagement lifetimes with survival $\sim t^{-\beta}$ the rate spectrum obeys:
$p(\lambda)\sim C_0\,\lambda^{\beta-1}\quad (\lambda\to 0^+),\qquad 0<\beta<1$. Keeping the leading term in \eqref{eq:phigen} and absorbing bounded prefactors into a constant $K$, we obtain:
\begin{equation}
C_a(\tau)\;\sim\;K\int_0^\infty \lambda^{\beta-1}\,e^{-c\,\lambda|\tau|}\,d\lambda.
\label{eq:S3}
\end{equation}
This integral can be evaluated explicitly by the changing variables
$u=c\,\lambda|\tau|$:
\[
\int_0^\infty \lambda^{\beta-1}e^{-c\,\lambda|\tau|}d\lambda
=(c|\tau|)^{-\beta}\int_0^\infty u^{\beta-1}e^{-u}du
=(c|\tau|)^{-\beta}\,\Gamma(\beta).
\]
Hence
\begin{equation}
C_a(\tau)\;\sim\;\tilde K\,|\tau|^{-\beta},\qquad
\tilde K=K\,c^{-\beta}\Gamma(\beta),\quad 0<\beta<1,
\label{eq:S4}
\end{equation}
\emph{independently of whether the single engagement pulse decays from $t=0$ or builds and then ends abruptly}. The only ingredient that matters for the tail is the
small $\lambda$ behavior of $p(\lambda)$.

Consequently, using $C_a(\tau)\sim |\tau|^{\alpha}$ with $\alpha=-\beta$, the MSD scales as (as shown in Section~S2):
\[
\langle x^{2}(t)\rangle \propto t^{\,2+\alpha},
\]
with the special case $\beta=\tfrac{1}{2}$ (i.e., $\alpha=-\tfrac{1}{2}$) giving $\langle x^{2}(t)\rangle \propto t^{3/2}$.

\medskip
\noindent\textbf{Summary.}
Modeling the engaged interval as a build then abrupt "off" pulse, and taking the acceleration as its time derivative, yields a single-pulse autocorrelation
$\phi_{\lambda,T}(\tau)$ with the same long-time exponential factor $e^{-\lambda|\tau|}$ as the
OU kernel (Eq.~\eqref{eq:phicut}). Mixing over a spectrum of rates with
$p(\lambda)\sim \lambda^{\beta-1}$ produces the power-law tail \eqref{eq:S4},
$C_a(\tau)\sim |\tau|^{-\beta}$, and the MSD scaling $\langle x^2(t)\rangle\sim t^{2-\beta}$,
irrespective of whether a single pulse \emph{decays} from $t=0$ or \emph{builds and ends abruptly}.
Thus the long-time exponent is determined by the rate spectrum (lifetime statistics), not by the micro-shape of traction within an engagement.

The analysis in this section assumes that the autocorrelation associated with a single engaged interval is short-ranged, with a decay scale set by the engagement lifetime $1/\lambda$. This is expected for clutch-mediated traction because adhesion turnover and detachment terminate individual traction engagement episodes and impose a finite-memory cutoff. If, instead, a single event impulse had intrinsic long-memory power law decay (in the absence of a detachment cutoff), that intrinsic tail could dominate the mixture integral and modify the asymptotic form of $C_a (\tau)$. In the experimentally relevant regime considered here, the long-time scaling is therefore controlled by the small-$\lambda$ sector of $p(\lambda)$, not by the detailed pulse micro-shape.

\section{From ballistic-pausing motion of single axons to superdiffusive ensemble dynamics}

In Sec.~S4 we showed that if the acceleration autocorrelation decays as $C_a(\tau)\sim |\tau|^{\alpha}$ with $\alpha=-\beta$ for some $0<\beta<1$, then the population MSD obeys $\langle x^2(t)\rangle\propto t^{2+\alpha}$, independent of the detailed build/cutoff shape of traction pulses (Eqs.\ ~\eqref{eq:S17}--~\eqref{eq:S4}). 
Here we demonstrate, by an explicit two-state construction, that \emph{ballistic} single-axon motion interleaved with \emph{pauses} yield the same long-time exponents at the ensemble level, without implying any gradual slowing within a single run.

\medskip
\noindent\textbf{Model.} Consider a single axon alternating between two motility states: \emph{advance} at constant speed $v_0>0$ and \emph{pause} at speed $0$.
Let $\sigma(t)\in\{0,1\}$ indicate the motility state: $\sigma(t)=1$ during advance and $\sigma(t)=0$ during pause. 
Define the velocity and position as
\begin{equation}
v(t)=v_0\,\sigma(t),\qquad x(t)=\int_0^t v(s)\,ds.
\label{eq:S23_def}
\end{equation}

We assume that advance and pause dwell times alternate and are identically distributed in their respective states, with probability density functions (PDFs) $\psi_{\rm on}(t)$ and $\psi_{\rm off}(t)$ and heavy-tailed survival functions:

\begin{equation}
\Psi_{\rm on}(t)=\int_t^\infty \psi_{\rm on}(u)\,du \sim c_{\rm on}\,t^{-\beta_{\rm on}},
\qquad
\Psi_{\rm off}(t)\sim c_{\rm off}\,t^{-\beta_{\rm off}},
\qquad 0<\beta_{\rm on},\beta_{\rm off}<1,
\label{eq:S24}
\end{equation}
on an intermediate asymptotic range (physical cutoffs are implicit).
We write $\beta=\min\{\beta_{\rm on},\beta_{\rm off}\}$.

\medskip
\noindent\textit{Velocity autocorrelation as a same-state probability.}
Since $v(t)=v_0\,\sigma(t)$, one has
\begin{equation}
C_v(\tau)=v_0^{\,2}\,\langle \sigma(t)\,\sigma(t+\tau)\rangle.
\label{eq:S25}
\end{equation}
The factor $\langle \sigma(t)\,\sigma(t+\tau)\rangle$ is precisely the probability that the process is in the \emph{advance} state at both times $t$ and $t+\tau$.
We can decompose this probability according to whether there is a switch in $(t,t+\tau)$.
If at time $t$ the process is in advance, then the event “no switch occurs up to lag $\tau$’’ is the event that the residual time $T_{\rm on}$ left in the current advance interval exceeds $\tau$.
For alternating renewals, conditioning on being in a given state at time $t$ induces a duration-biased interval distribution. Consequently the residual-time tail satisfies (for large $\tau$ in the scaling range):
\begin{equation}
P(T_{\rm on}>\tau) \sim \tilde c_{\rm on}\,\tau^{-\beta_{\rm on}},
\qquad
P(T_{\rm off}>\tau) \sim \tilde c_{\rm off}\,\tau^{-\beta_{\rm off}},
\label{eq:S26}
\end{equation}
with positive constants $\tilde c_{\rm on/off}$. Here $P(A)$ denotes the probability for an event $A$. Eq.~\eqref{eq:S26} follows directly from the identity
\[
P(T>\tau)=\frac{\int_{\tau}^{\infty}\Psi(u)\,du}{\int_0^{\infty}\Psi(u)\,du},
\]
valid for the heavy-tail form $\Psi(u)\sim c\,u^{-\beta}$ ($0<\beta<1$), which yields $\int_\tau^{\infty}\Psi(u)\,du \propto \tau^{\,1-\beta}$ and hence $P(T>\tau)\propto \tau^{-\beta}$ after normalization.
Contributions that involve two or more switches in $(t,t+\tau)$ are of higher order in $\tau$ (they require successive short intervals) and do not affect the leading power.

Collecting the leading contributions from starting in advance and from starting in pause, one arrives at
\begin{equation}
 \ C_v(\tau) \;\sim\; K\,\tau^{-\beta}, \qquad 0<\beta<1, 
\label{eq:S27}
\end{equation}
with a constant prefactor $K>0$ that depends on $v_0$ and on the relative occupation of the two states but not on $\tau$.
Equivalently, with the notation used in the main text, $\alpha=-\beta$ and $C_v(\tau)\sim \tau^{\alpha}$.
This is the same exponent inferred there from the colored-acceleration (rate-spectrum) description: the heavy tail in the dwell-time statistics of advance/pause states and the heavy tail in the small–$\lambda$ regime of the rate spectrum $p(\lambda)\sim \lambda^{\beta-1}$ are two coarse-grained routes to the same asymptotic power law.

\begin{table}[h]
\centering
\begin{ruledtabular}
\begin{tabular}{cccccc}
$\beta_{\rm on}$ & $\beta_{\rm off}$ & $\min\{\beta_{\rm on},\beta_{\rm off}\}$ & $\alpha$ & MSD exponent $\,2+\alpha\,$ & Comment \\
\hline
0.50 & 0.50 & 0.50 & $-0.50$ & $1.50$ & Active-polymer FPT benchmark \\
0.50 & 0.70 & 0.50 & $-0.50$ & $1.50$ & On-state controls the tail \\
0.60 & 0.60 & 0.60 & $-0.60$ & $1.40$ & Symmetric tails \\
0.60 & 0.80 & 0.60 & $-0.60$ & $1.40$ & On-state controls the tail \\
0.80 & 0.60 & 0.60 & $-0.60$ & $1.40$ & Off-state controls the tail \\
0.70 & 0.90 & 0.70 & $-0.70$ & $1.30$ & Heavier tail $\Rightarrow$ smaller MSD exponent \\
0.90 & 0.90 & 0.90 & $-0.90$ & $1.10$ & Near-diffusive crossover window \\
\hline
\multicolumn{2}{c}{Experiment} & $\approx 0.60$ & $-0.60$ & $1.40$ & Matches Figs.~2–3
\end{tabular}
\end{ruledtabular}
\caption{\label{tab:model}
Predicted ensemble exponents from ballistic–pausing kinetics with heavy-tailed dwell times.
For given tail exponents $(\beta_{\rm on},\beta_{\rm off})$, the velocity-autocorrelation exponent is
$\alpha=-\min\{\beta_{\rm on},\beta_{\rm off}\}$ and the MSD exponent is
$2+\alpha$ (see Eq.~\eqref{eq:S27} and Sec.~S2).
The last row (“Experiment”) highlights the measured velocity autocorrelation decay $C_v(\tau)\propto \tau^{-0.6}$,
which implies $\beta\simeq 0.6$ and thus $\langle x^2(t)\rangle\propto t^{1.4}$, consistent with
Figs.~2–3 of the main text.}
\end{table}

Table~\ref{tab:model} summarizes, for representative values of the dwell-time tail exponents $(\beta_{\rm on},\beta_{\rm off})$, the predicted velocity-autocorrelation exponent $\alpha=-\min\{\beta_{\rm on},\beta_{\rm off}\}$ and the corresponding ensemble MSD exponent $2+\alpha=2-\min\{\beta_{\rm on},\beta_{\rm off}\}$ quoted in the main text.
The entries are obtained by inserting $(\beta_{\rm on},\beta_{\rm off})$ into Eq.\ \eqref{eq:S27} for $C_v$ and into the MSD relation established earlier in Sec.~S2, using the measured velocity autocorrelation exponent as a check.
In particular, the experimentally observed velocity autocorrelation function decay $C_v(\tau)\propto \tau^{-0.6}$ corresponds to $\beta\simeq 0.6$ and implies the MSD scaling $\langle x^2(t)\rangle\propto t^{1.4}$, in agreement with Figs.\ 2–3 of the main text.

\medskip
\noindent\textbf{Physical picture.}
Within each advance interval the speed is approximately constant ($v_0$). There is no gradual deceleration during a single run.
Long-range temporal memory arises because the durations of advance and pause intervals are broadly distributed: the probability that two runs separated by $\tau$ fall in the same motility state decays as a power law $\tau^{-\beta}$.
Since $v(t)$ equals $v_0$ in advance and $0$ in pause, this same-state probability directly sets $C_v(\tau)$ up to a constant factor (Eq.\ \eqref{eq:S25}).
Thus constant-velocity runs separated by pauses naturally yield the measured ensemble power laws without invoking any monotonic slowing of individual growth cones.
Quantitatively, our data give $\alpha\simeq -0.6$ for the velocity autocorrelation function, consistent with $\beta\simeq 0.6$ in Eq.\ \eqref{eq:S27} and with the MSD exponent $1.4$ reported in the main text. We note that the expression $C_v(\tau)=\langle v(t)\,v(t+\tau)\rangle$ agrees with the usage in Sec.~S3. Rigorously, one may interpret the average as an ensemble average or as a long-time average in the stationary renewal limit at fixed time lag $\tau$.
If a mean subtraction is considered, one replaces $C_v$ by the velocity covariance. All asymptotic exponents stated above are unchanged.

The two state advance–pause construction in this section is included as a constructive illustration of the same long-memory mechanism encoded by the small-$\lambda$ tail of the rate spectrum in Sec.~4. We emphasize that this model is not a distinct microscopic alternative to the MaxEnt/OU-mixture framework, and it does not imply a unique description at the single-axon level. Rather, it shows that widely observed single-cone dynamics (constant speed runs separated by pauses) can produce the same ensemble power laws when run and pause durations are broadly distributed. In this sense, the results in this section support the universality of the scaling exponents: multiple coarse-grained representations that share the same heavy-tailed lifetime statistics lead to the same long-time behavior for the MSD and the velocity autocorrelation function.

\section{Statistical comparison with diffusive and ballistic MSD scaling}

We performed a statistical test of whether the long-time mean squared
displacement (MSD) data in Fig.~4(a) are consistent with a linear-in-time (diffusive)
growth law or with a simple drift-dominated (ballistic) growth law. Since the data
identifies a crossover near $t\simeq 30~\mathrm{h}$, we tested the long-time
regime $t\ge 30~\mathrm{h}$, using the data
\begin{equation}
\begin{split}
(t_i,M_i)=&
(30,2220),(40,3170),(50,4501),(60,5580),\\
& (70,7190),(80,8520),(90,10195),
\end{split}
\end{equation}
where $M_i$ denotes the measured MSD. The colored-noise model predicts a power-law MSD of the form
\begin{equation}
  M(t)\equiv \langle \Delta x^2(t)\rangle \propto t^\nu,
\end{equation}
with $\nu=2+\alpha$. We therefore fit the logarithmic model
\begin{equation}
  \log M_i = a + \nu \log t_i + \epsilon_i ,
  \label{eq:S_log_fit}
\end{equation}
and perform a regression $t$-test for the slope in the log--log fit.

First, we compare the data with a linear-in-time MSD baseline. This corresponds
to the null hypothesis $H_0:\nu=1$, whereas superdiffusive MSD growth corresponds to
$H_1:\nu>1$.

An ordinary least-squares fit of Eq.~\eqref{eq:S_log_fit} gives
\begin{equation}
  \hat{\nu}=1.395,
  \qquad
  \mathrm{SE}(\hat{\nu})=0.022 .
\end{equation}
The corresponding test statistic is
\begin{equation}
  t=\frac{\hat{\nu}-1}{\mathrm{SE}(\hat{\nu})}
   =\frac{1.395-1}{0.022}
   =18.1 ,
\end{equation}
with $n-2=5$ degrees of freedom, giving the corresponding $p$-value
\begin{equation}
  p=4.7\times 10^{-6}.
\end{equation}
The 95\% confidence interval for the MSD scaling exponent is
\begin{equation}
  \nu = 1.395 \pm 0.056,
  \qquad
  1.339 \le \nu \le 1.451 .
\end{equation}
Thus the fitted MSD exponent is statistically distinguishable from unity in
the long-time regime, supporting superdiffusive rather than linear-in-time MSD
growth.

We also compare the data with a simple drift-dominated model for axonal
extension. If the axonal displacement grew by ordinary linear drift,
$L(t)\propto t$, then the corresponding MSD would scale ballistically,
\begin{equation}
  M(t)=L^2(t)\propto t^2 .
\end{equation}
This gives the null hypothesis
\begin{equation}
  H_0:\nu=2,
\end{equation}
with the alternative
\begin{equation}
  H_1:\nu<2.
\end{equation}
Using the same fit, the test statistic is
\begin{equation}
  t=\frac{\hat{\nu}-2}{\mathrm{SE}(\hat{\nu})}
   =\frac{1.395-2}{0.022}
   =-27.8 ,
\end{equation}
again with $n-2=5$ degrees of freedom. The corresponding $p$-value is
\begin{equation}
  p=5.7\times 10^{-7}.
\end{equation}
The confidence interval above also excludes $\nu=2$. Therefore, the long-time
MSD data are not consistent with a simple drift-dominated linear-advance model,
which would produce ballistic MSD scaling.

As an additional check, we tested for curvature in the MSD without assuming a
power-law form by comparing the nested models
\begin{equation}
  H_0: M(t)=a+bt
\end{equation}
and
\begin{equation}
  H_1: M(t)=a+bt+ct^2 .
\end{equation}
For the long-time data, the fitted quadratic coefficient is positive,
$c=0.557$, and the extra sum-of-squares test gives
\begin{equation}
  F=24.6,
  \qquad
  p=7.7\times 10^{-3}.
\end{equation}
This independent curvature test indicates a statistically significant
departure from a purely linear-in-time MSD baseline in the post-crossover
regime.

Together, these tests show that the long-time MSD data are not statistically
consistent with either a linear-in-time MSD baseline or a simple
drift-dominated ballistic scaling law. Instead, the fitted exponent lies in
the superdiffusive range and is close to the scaling expected from the
colored-noise model, for which
\begin{equation}
  \nu = 2+\alpha .
\end{equation}
For the experimentally relevant value $\alpha\simeq -0.6$, this gives
$\nu\simeq 1.4$, in agreement with the fitted value
$\hat{\nu}=1.395$.

\end{document}